\documentclass[prd,a4paper,10pt,showpacs,preprint,byrevtex,twocolumn]{revtex4} %

\usepackage{amsfonts}
\usepackage{bm,amssymb}
\usepackage{amsmath}

\begin{document}

\title{Spin dependent operators in correlated gaussian bases}

\author{Bernard \surname{Silvestre-Brac}}
\email[E-mail: ]{silvestre@lpsc.in2p3.fr} \affiliation{LPSC,
Universit\'{e} Joseph Fourier Grenoble 1, CNRS/IN2P3, Institut National
Polytechnique de Grenoble, 53 Avenue des Martyrs, F-38026 Grenoble-Cedex,
France}
\author{Vincent \surname{Mathieu}}
\email[E-mail: ]{vincent.mathieu@umh.ac.be} \affiliation{Groupe de
Physique Nucl\'{e}aire Th\'{e}orique, Universit\'{e} de
Mons-Hainaut, Acad\'{e}mie universitaire Wallonie-Bruxelles, Place
du Parc 20, B-7000 Mons, Belgium}

\date{\today}

\begin{abstract}
In their textbook, Suzuki and Varga [Y. Suzuki and K. Varga, {\em
Stochastic Variational Approach to Quantum-Mechanical Few-Body
Problems} (Springer, Berlin, 1998)] present the stochastic
variational method with the correlated Gaussian basis in a very
exhaustive way. The matrix elements for central potentials are put
under a pleasant form but the elements for spin dependent operators,
when treated, are given as very cumbersome expressions. In this paper,
we find a lot of new formulae for those elements. Their expressions
are given in terms of the same geometrical functions that appear in
the case of central potentials. These functions get therefore a
universal status; this property is very useful for numerical
applications.
\end{abstract}


\maketitle

\section{Introduction}
\label{sec:intro}
There exist several different technical methods to solve the few-body problem
with accuracy: Monte Carlo calculations \cite{ham94}, Faddeev and Yakubovsky
treatments \cite{glo96}, hyperspherical formalism \cite{fab83}, expansion on
various types of orthogonal \cite{bsb93}, or non orthogonal bases \cite{suzu98}.
Each technics shows specific advantages and drawbacks.
Among others, the stochastic variational method is especially attractive. It
relies on expansion of the wave function in term of gaussian type functions.
The stochastic algorithm allows to consider very large bases with a minimum of
variational effort. The drawback of this method is the non orthogonality of
basis wave functions with the possibility of appearance of spurious states due
to overcompleteness; if this last inconvenience is overcome, using non orthogonal
bases is not really a problem. The generalized eigenvalue problem arising in
this case is well under control nowadays. The great advantage of using gaussian
type functions is the rapid convergence and, above all, the possibility to
compute the resulting matrix elements with analytical expressions most of
time.

The stochastic variational method is described in full details in the remarkable
textbook by Y. Suzuki and K. Varga \cite{suzu98}, where most important and
fundamental formulae are derived. This very complete work will be refered as SV
throughout this paper, and all subsequent references can be found in it.

However a number of important formulae are missing and do not seem
to appear in the litterature. This was the case of the Fourier
transform of the general correlated Gaussian and its application to
the matrix elements of the semi-relativistic kinetic energy
operator. We presented such corresponding formulae in a recent paper
\cite{bsbm07}, hereafter denoted SBM. In the same paper, we also
propose new formulae for the matrix elements of the central
potential, which are more efficient on the numerical point of view.
Moreover, in the SV book, the matrix elements of spin-dependent
operators are presented in a very cumbersome form.

Despite the fact that the calculations are more involved
than the simpler central potential case, we have found a formulation that allows
to express all these elements in an elegant and unified way. The aim of this
paper is to present these new formulae and to convince the reader that
complicated physical situations needing the use of central plus spin-dependent
operators can be treated in a unified way based on universal functions. The
gain of performance in a numerical treatment is very important.

We will derive expressions for the most general correlated gaussians
(arbitrary number of particles $N+1$, arbitrary angular momentum
$L$, arbitrary radial $K$ quantum number). We focus our interest on
spin-dependent operators, but relegate in appendix some
considerations on central potentials already given in SV and SBM, in
order to achieve some self-consistency. Moreover, the non-natural
parity states are very difficult to handle in correlated bases and,
in the following, we just study natural parity (i.e spatial parity
equal to $(-1)^L$) states.

To calculate the same matrix elements, we will propose four different
alternative expressions, with their own interests and drawbacks; this is very
useful for numerical checks. A first approach sticks closely to the spirit of
SV; the corresponding formulae exhibits nicely the symmetry properties. A
second approach, proposed in SBM, gives formulae in which the symmetry properties
are less transparent, but which are more efficient numerically. In both
approaches, we give a formulation in terms of potential integrals $J$ which
depend on combinations of Hermite polynomials and make economical the final
expression, and a formulation in terms of potential integrals $\mathcal{F}$
which are universal and, most of the time, analytical.

The paper is organized as follows. We recall, in a first section,
the description of the systems under consideration (intrinsic
coordinates, definition of correlated Gaussians and their generating
functions) and important definitions that are a link on all
expressions presented here. The third section is devoted to the
results concerning the spin-orbit operators, while the fourth
section deals with tensor operators. In the appendixes, we give a
number of important ingredients that should be used in the course of
the various derivations of this work. Most of them are rather
technical, but are also new relations that do not appear in the
literature.

\section{The system under consideration}
\label{sec:syst} Since a lot of details concerning the system under
consideration are already given in SBM, we recall here only a few
things, referring the reader to this work for further information.

\subsection{Jacobi coordinates}
\label{ssec:Jaccoord}

Let us denote by $N+1$ the number of particles ($N \geq 1$); the position of
particle $i$, of mass $m_i$, is $\bm r_i$ in some frame, while the conjugate
momentum is $\bm p_i$. In quantum mechanics and in position representation,
$\bm r_i$ and $\bm p_i = -i \partial/\partial \bm r_i$ are operators in the
Hilbert space of the states. The intrinsic properties of the system are
described in terms of $N$ Jacobi coordinates $\bm x_i$ while the bulk
properties are dependent of the center of mass coordinate $\bm R$.

In order to simplify the notations, let us introduce a ``super-vector''
$\tilde{\bm x} = (\bm{x}_1,\bm{x}_2,\ldots,\bm{x}_N)$ and write the
coefficients of linear combinations as a ``line (or column) matrix'', i.e
$\tilde{u} = (u_1,u_2,\ldots,u_n)$. This allows to shorten the expressions
using the usual matrix operations. For example, a linear combination of
Jacobi coordinates is denoted $\tilde{u} \bm x$ = $u_1 \bm x_1 + u_2 \bm x_2 +
\ldots u_N \bm x_N$. $A$ being a $N \times N$ matrix, the expression $A \bm x$
means the super vector $(\sum_j A_{1j} \bm x_j, \sum_j A_{2j} \bm x_j, \ldots,
\sum_j A_{Nj} \bm x_j)$. Lastly $\tilde{\bm x} \cdot \bm y$ represents the
scalar $\bm x_1 \cdot \bm y_1 + \bm x_2 \cdot \bm y_2 + \ldots + \bm x_N \cdot
\bm y_N$ where the symbol $\cdot$ deals with a spatial scalar product.

It is easy to show that both the operators $\bm r_i - \bm R$ appearing in one-body
potentials and the operators $\bm r_{ij}=\bm r_i - \bm r_j$ appearing in two-body
potentials are combinations of Jacobi coordinates. Explicitly
\begin{equation}
\label{eq:comblinpos}
\bm r_i - \bm R =  \tilde{w}^{(i)} \bm x \, ; \,
\bm r_{ij} = \bm r_i - \bm r_j = \tilde{w}^{(ij)} \bm x
\end{equation}
where the coefficients $\tilde{w}^{(i)}$ and $\tilde{w}^{(ij)}$ are mass
dependent.

The conjugate momenta of the Jacobi coordinates are denoted
$\tilde{\bm \pi}=(\bm \pi_1,\bm \pi_2,\ldots,\bm \pi_N)$ ($\bm \pi_i = -i
\partial/\partial \bm x_i$). In the same way, both the operator $\bm p_i $
(in the center of mass frame) appearing in one-body potentials and the operator
$\bm p_{ij} = (m_j \bm p_i - m_i \bm p_j)/(m_i+m_j)$ appearing in two-body
potentials are combinations of Jacobi momenta. Explicitly
\begin{equation}
\label{eq:comblinmom}
\bm p_i =  \tilde{\zeta}^{(i)} \bm \pi \, ; \,
\bm p_{ij} = (m_j \bm p_i - m_i \bm p_j)/(m_i+m_j)=\tilde{\zeta}^{(ij)} \bm \pi
\end{equation}
where, again, the coefficients $\tilde{\zeta}^{(i)}$ and $\tilde{\zeta}^{(ij)}$
are mass dependent.

In many variational methods, the wave function of the system is expanded on basis
states as
\begin{equation}
\label{eq:fotot}
\left|\Psi^{J M} \right\rangle=  \sum_i C_i \left|\Psi^{J M}_i\right\rangle.
\end{equation}

We forget about the colour, isospin and center of mass degrees of freedom and
consider only space and spin degrees of freedom, since we are concerned only
with spin-dependent potentials. $JM$ are the total spin and magnetic quantum
numbers for the system and $\left| \Psi^{J M}_i \right\rangle$ are basis trial
wave functions.

If some of the particles are identical, the basis states must be
(anti)symmetrized with the help of a (anti)symmetrizer operator. In the
stochastic variational method the resulting complications are not really a
problem because correlated Gaussians with permuted coordinates are again
correlated Gaussians with modified parameters and the spin functions can be
handled with the well controled Racah algebra. Thus the (anti)symmetrization
complication essentially results only in adding linear combinations of wave
functions of the same type. Thus, in order to simplify the presentation, we
suppose basis wave functions that are not (anti)symmetrized.

In order to exploit fully the power of the formulation in terms of correlated
Gaussians, one must adopt a coupling for angular momenta that is of type $LS$
with a spatial wave function with total angular momentum $L$ and a spin wave
function of total spin $S$.

With this in mind, one writes one basis state as
\begin{equation}
\label{eq:basisst}
\left| \Psi^{J M}_i \right\rangle = \left[\left| \Psi_{L_i}(1,\ldots,N+1)
\right\rangle \left| \chi_{S_i}(1,\ldots,N+1)\right\rangle\right]_{JM}.
\end{equation}

Concerning the spin function, there may occur several possibilities for the
coupling to a spin $S_i$. In general, various intermediate couplings appear
and are part of the quantum numbers necessary to label a basis state $i$.
For example, in the three-body problem, one chooses the coupling
$[(s_1,s_2)_{S_{12i}}s_3]_{S_i}$ with only one intermediate quantum number
$S_{12i}$. In contrast, in the four-body problem, one may choose the
coupling $\left[\left((s_1 s_2)_{S_{12i}}s_3\right)_{S_{123i}} s_4\right]_{S_i}$
with the two intermediate quantum numbers $(S_{12i},S_{123i})$ or the
coupling $\left[\left(s_1 s_2\right)_{S_{12i}}\left(s_3 s_4 \right)_{S_{34i}}
\right]_{S_i}$ with the two intermediate quantum numbers $(S_{12i},S_{34i})$.

The spatial wave function is discussed more deeply in the next subsection.

\subsection{Correlated Gaussians}
\label{subsec:corgaus}

The so-called ``correlated Gaussian'' is a special form of space wave function
which is widely used in the stochastic variational method. It is expressed in
terms of the Jacobi coordinates. It has a number of advantages. In its most
general version the convergence in terms of basis states is quite fast, and
the way to deal with the angular momentum is in a form that allows to treat
easily systems with arbitrary number of particles. Moreover, one can obtain
most of the matrix elements under an analytical expression.

The argument of the exponential is a bilinear combination of the
Jacobi coordinates: $\sum_{i,j=1}^{N} A_{ij} \bm{x}_i \cdot
\bm{x}_j$ = $\tilde{\bm{x}} \cdot A \bm{x}$. The matrix $A$ must be
symmetric ($A =\tilde{A}$) and positive definite.

To deal with a non vanishing total angular momentum, one must introduce
spherical harmonics somehow or other. The most elegant manner is to use a
\textbf{single} solid harmonic $\mathcal{Y}_{LM}(\bm v) = v^L Y_{LM}(\hat{v})$.
To achieve some symmetry, and also to have more variational parameters at our
disposal, the argument of the solid harmonic is the most general linear
combination of the Jacobi coordinates $\bm v = \sum_{i=1}^N u_i \bm{x}_i$.

With those definitions, the most general correlated Gaussian is given by
(note a slight difference with SV notations; their matrix $A$ is twice ours and
moreover $N$ is the number of Jacobi coordinates while SV consider it as the
number of particles)
\begin{eqnarray}
\label{eq:corgaus}
&\left\langle \bm x \vert \Psi_{KLM}(u,A) \right\rangle = f_{KLM}(u,A;\bm x) =&
\nonumber \\
&\exp(- \tilde{\bm x} \cdot A \bm x)\:|\tilde{u} \bm x|^{2K}
\mathcal{Y}_{LM}(\tilde{u} \bm x).&
\end{eqnarray}

Thus, each basis state is described by $N(N+3)/2$ free parameters ($N(N+1)/2$
for the matrix $A$ and $N$ for the vector $u$). This prescription
(\ref{eq:corgaus}) is only able to deal with natural parity states. The term
$|\tilde{u} \bm x|^{2K}$ is introduced for generality and to treat with more
accuracy potentials with specific singular features. However, it complicates
a lot the resulting expressions. It is often more convenient (except when
the potential is so singular that the resulting integrals diverge) to keep in
the calculation the correlated Gaussians restricted to $K = 0$, including more
basis states to compensate a slower convergence.

\subsection{Matrix elements and generating functions}
\label{subsec:genfunc}
As explained in SV and SBM, the calculation of the matrix elements of some
operator $\hat{O}$ on the correlated gaussians, namely $\langle
\Psi_{K'L'M'}(u',A')| \hat{O} | \Psi_{KLM}(u,A) \rangle$ relies on the
generating function technics.

Let us define the functions
\begin{equation}
\label{eq:genfunc}
g(\bm s,A; \bm x) = \exp(- \tilde{\bm x} \cdot A \bm x + \tilde{\bm s} \cdot
\bm x)
\end{equation}
where $\bm s$ is an arbitrary super-vector, $\tilde{\bm s} = (\bm s_1,\bm s_2,
\ldots,\bm s_N)$.

The $g$ functions are called the generating functions for the
correlated Gaussians since one has
\begin{equation}
\label{eq:fvsgen}
\begin{split}
 f_{KLM}(u,A;\bm x) = &  \frac{1}{B_{KL}}\\
         \times \int d\hat{\bm{e}} \:  Y_{LM}(\hat{\bm{e}}) & \left(
\frac{\partial^{2K+L}}{\partial \lambda^{2K+L}} g(\lambda \bm e
u,A; \bm x)\right)_{\lambda=0, |\bm e| = 1}
\end{split}
\end{equation}
where the geometrical coefficient $B_{KL}$ is defined as
\begin{equation}
\label{eq:BKL}
B_{KL}= \frac{4 \pi (2K+L)!}{2^K \, K! \, (2K+2L+1)!!}.
\end{equation}
In Eq. (\ref{eq:fvsgen}), the super-vector $\bm s = \lambda \bm e u$ must
be understood with all its components proportional to the same three vector
$\bm e$, namely $\bm s_i = \lambda u_i \bm e$.

Using Eq. (\ref{eq:fvsgen}) in the expression of the searched matrix element
leads to
\begin{equation}\label{eq:elmatgen}
\begin{split}
\langle \Psi_{K'L'M'}(u',A')| \hat{O} |\Psi_{KLM}(u,A)\rangle   \\
= \frac{1}{B_{K'L'}B_{KL}}\int d\hat{\bm{e}}\:d\hat{\bm{e}}'
Y_{LM}(\hat{\bm{e}}) Y_{L'M'}^\ast(\hat{\bm{e}}')
\\
\times\left( \frac{\partial^{2K'+L'+2K+L}}{\partial
{\lambda'}^{2K'+L'}
\partial \lambda^{2K+L}} \left\langle {\cal O}\right\rangle  \right)_{\lambda =
\lambda' =0,|\bm e| = |{\bm e}'| = 1},
\end{split}
\end{equation}
with the matrix element between the generating functions
\begin{equation}\label{}
\left\langle {\cal O}\right\rangle =\langle g(\lambda' {\bm e}'
u',A';\bm x) | \hat{O} | g(\lambda \bm e u,A;\bm x)\rangle.
\end{equation}
The matrix element that is left for computation is now between the generating
functions, the form of which is much simpler.

Whatever the operator used in the Hamiltonian, it is scalar for
rotations. Thus, the matrix elements do not depend on the magnetic
quantum number $M$. Central potentials are spin independent; they
have been treated extensively in BSM; some interesting formulae are
gathered in appendix \ref{sec:formconnu}. Here, we are mainly
concerned with spin-orbit operators and tensor operators. Both are
spin dependent.

\subsection{Universal functions}
\label{subsec:univfunc}
Generally, the expression for the matrix elements needs the introduction of
several types of quantities:
\begin{itemize}
    \item
Dynamical quantities are functions of the explicit form of the
operator $\hat{O}$. Any potential depends on some function of the Jacobi
variables, for example for a two-body potential : $V(|\bm r_i - \bm r_j |)$
= $V(|\tilde{w}^{(ij)} \bm x|)$ (see (\ref{eq:comblinpos})). In SV the
dynamical resulting quantity is an integral of type
\begin{equation}
\label{eq:intJnc}
J(n,c)=\frac{1}{\sqrt{\pi}} \int_0^\infty V(x\sqrt{2 /c}) e^{-x^2} Q_n(x) \, dx
\end{equation}
where $Q_n(x)$ is a specific function expressed in terms of Hermite polynomials.
For the case of central potentials $Q_n(x)=H_1(x)H_{2n+1}(x)/(2n+1)!$.

In BSM, we proposed a new formulation for the matrix elements, which needs a
more general integral
\begin{equation}
\label{eq:intJnac} J(n,\alpha,c)=\frac{1}{\sqrt{\pi}}
\int\limits_0^\infty V\left(x\sqrt{2\frac{\alpha}{c}}\right)
e^{-\alpha x^2} Q_n(x) \, dx,
\end{equation}
with the same $Q_n(x)$ function. Obviously $J(n,c) = J(n,\alpha=1,c)$. The
reason for appearance of such integrals is explained in Appendix
\ref{sec:specexp}.

Although these functions are the most economical for a general
presentation, they have several drawbacks: The function $Q_n(x)$
depends on each type of potential considered (they are not identical
for central, spin-orbit and tensor interactions), and the analytical
expression of the $J$ integrals, even for the simplest forms of
$V(x)$, is not very simple. This is why, in BSM, we proposed also
alternative expressions for the matrix elements in terms of a
universal integral
\begin{equation}
\label{eq:intFV} \mathcal{F}_V(k,A) = \int_0^\infty V(u) u^k
e^{-Au^2} \: du,
\end{equation}
with integer value for $k$. Indeed, a lot of closed expressions exist for
various forms of potentials $V(u)$.

\item
Pure geometrical quantities also appear in the formalism. For example, the
coefficients $B_{KL}$ are very common (see (\ref{eq:BKL})). In the
same way, the coefficients $Z_l^\lambda$, as defined by (\ref{eq:coefZ}), are
also of some use.

\item
Lastly, geometrical functions depending on the various parameters of the
problems were introduced both in SV and BSM.
In SV, a very important function, appearing in the case of central potential,
is defined by
\begin{equation}\label{eq:fFVarg}
\begin{split}
F^n_{p,p',l}(u,u',v,w,w')=n! \sum_{m=0}^{p} \sum_{m'=0}^{p'}
\frac{u^{p-m}}{(p-m)!} \frac{{u'}^{p'-m'}} {(p'-m')!}\\ \times
\frac{v^{l-n+m+m'}}{(l-n+m+m')!}\frac{w^{n+m-m'} {w'}^{n-m+m'}}
{2^{m+m'}\,m!\,m'!(n-m-m')!}
\end{split}
\end{equation}
(we have also the following constraints $p+p'+l \geq n$ and $n - l \leq m+m'
\leq n$).

In BSM, another function of great importance is defined by
\begin{equation}
\label{eq:fFmoi}
\begin{split}
 &F^{K,K',L}_{n,k}(x,x',y,y')=n! \sum_{m=\max(k+L,n-K')}^{\min(n-k,K+L)}
 \frac{x^{K+L-m}}{(K+L-m)!}\\
 &\times \frac{{x'}^{K'-n+m}}{(K'-n+m)!} \frac{y^{2m-L}{y'}^{2(n-m)+L}}
{(m-k-L)!(n-k-m)!}.
\end{split}
\end{equation}
The functions appearing in (\ref{eq:fFVarg}) and (\ref{eq:fFmoi}) are
denoted with the same letter $F$ but are not identical. They differ by the
number of continuous arguments. The formulation in terms of $F$ function
(\ref{eq:fFmoi}) is more efficient numerically since the number of arguments is
less (4 instead of 5) and the sum runs on a smaller number of indices (1 instead
of 2). For central potential matrix elements, the $F$ functions are
associated to the $J$ integrals.

Associated to the $\mathcal{F}$ integrals, other functions, more complicated,
are necessary. They are absent in SV, but have been given in BSM. In a first
formulation, we need the following function
\begin{equation}
\label{eq:fHnk}
\begin{split} &H^{K,K',L}_{n,k}(x,x',y)=
\\ &\sum_{r=0}^{K+K'+L-n} (-1)^r \frac{(K+K'+L-r)! \: y^r}
{(K+K'+L-n-r)!} G^{K,K',L}_{k,r}(x,x')
\end{split}
\end{equation}
and
\begin{equation}
\label{eq:fGkr}
\begin{split}
G^{K,K',L}_{k,r}(x,x')& = \sum_{s=0}^{K-k} \sum_{s'=0}^{K'-k}
\frac{x^s {x'}^{s'}}{s! (K-k-s)! s'! (K'-k-s')! } \\
&\times\frac{1}{(r-s-s')!(2k+L+s+s'-r)!}
\end{split}
\end{equation}
while in a second formulation, it appears the following function
\begin{eqnarray}
\label{eq:fPnk}
 & & P^{K,K',L}_{n,k}(x,x',y,y',z)= \\
 & &\sum_{r=k+L}^{K+L} \frac{x^{K+L-r} x'^{K'+r-n} y^{2r-L}
{y'}^{2(n-r)+L}} {(K+L-r)!(r-k-L)!} M^{K'}_{n,k,r}(z)
 \nonumber
\end{eqnarray}
with
\begin{equation}
\label{eq:fGnkr} M^{K'}_{n,k,r}(z) =\!\!\!\!\!\!\!\!\!\!\!\!\!\!\!
\sum_{s=\max(0,k+r-n)}^{K'+r-n} \!\!\!\!\!
\frac{(s+n)!}{(K'+r-s-n)!(s+n-k-r)!} \frac{z^s}{s!}.
\end{equation}

Here again, the second formulation (\ref{eq:fPnk}) is numerically
more efficient since the $P$ function requires a summation on 2 indices
while the $H$ function requires 3 indices.
\end{itemize}

As we will see, all the quantities presented in this section -- dynamical,
purely geometrical factors or geometrical functions -- have indeed a universal
character and appear not only for central potentials, but also for any
type of complicated potentials. This consequence is very beneficial on a
numerical point of view since it allows to treat very different types of
potentials with the same basic ingredients, which can be computed once and
for all.

\section{Spin-orbit potentials}
\label{sec:sporbpot}
In this section, we discuss about the spin-orbit potentials. Schematically, we
have three types of spin-orbit potentials.
\begin{itemize}
    \item One-body spin-orbit potential where the form of the operator for the
particle $i$ is given by (in the center of mass frame for the system)
\begin{equation}
\label{eq:sporb1c}
V_i = V( | \bm r_i|) \bm L_i \cdot \bm s_i
\end{equation}
where $\bm s_i$ is the spin of the particle located at $\bm r_i$, with an
angular momentum $\bm L_i = \bm r_i \times \bm p_i$, relative to the center of
mass. Expressed in terms of Jacobi coordinates, it writes
\begin{equation}
\label{eq:sporb1cjac}
V_i = V( | \tilde {w}^{(i)} \bm x|) (\tilde {w}^{(i)} \bm x \times
\tilde {\zeta}^{(i)} \bm \pi ) \cdot \bm s_i.
\end{equation}
This kind of potential appears for instance as a relativistic
correction of a flux tube model for hadron confining potential
\cite{Fabien}.

    \item What is called usually ``spin-orbit'' potential is the two-body
symmetric spin-orbit potential, whose form is
\begin{equation}
\label{eq:sporb2cs}
V_{ij} = V( | \bm r_i - \bm r_j|) \bm L_{ij} \cdot \bm S_{ij}
\end{equation}
where $\bm S_{ij} = \bm s_i + \bm s_j$ is the total spin of the pair $(i-j)$,
while $\bm L_{ij} = \bm r_{ij} \times  \bm p_{ij}$ is the angular momentum
of the pair in its center of mass frame. Expressed in terms of Jacobi coordinates,
it writes
\begin{equation}
\label{eq:sporb2csjac}
V_{ij} = V( | \tilde {w}^{(ij)} \bm x|) (\tilde {w}^{(ij)} \bm x \times
\tilde {\zeta}^{(ij)} \bm \pi ) \cdot \bm S_{ij}.
\end{equation}
This form of potential is traditional as a relativistic correction of one-photon,
one-boson or one-gluon exchange potentials.

    \item Sometimes, it is necessary to introduce the two-body antisymmetric
spin-orbit potential, defined by
\begin{equation}
\label{eq:sporb2ca}
V_{ij} = V( | \bm r_i - \bm r_j|) \bm L_{ij} \cdot \bm \Delta_{ij}
\end{equation}
and which is very similar to (\ref{eq:sporb2csjac}) but with the spin operator
$\bm \Delta_{ij} = \bm s_i - \bm s_j$. Expressed in terms of Jacobi coordinates,
it writes
\begin{equation}
\label{eq:sporb2cajac}
V_{ij} = V( | \tilde {w}^{(ij)} \bm x|) (\tilde {w}^{(ij)} \bm x \times
\tilde {\zeta}^{(ij)} \bm \pi ) \cdot \bm \Delta_{ij}.
\end{equation}
This potential is also a relativistic correction of one-photon, one-boson or
one-gluon exchange potentials. However, being proportional to $1/m_i^2 - 1/m_j^2$,
it has no effect in the case of identical particles.
\end{itemize}

In order to simplify the notation let us note any of these potentials as
$V \mathcal{L} \cdot \mathcal{S}$ with obvious identification. $\mathcal{L}$ and
$\mathcal{S}$ being vector operators, the matrix elements of the potential are
obtained with help of Wigner-Eckart (WE) formalism. Using (\ref{eq:elmatprodtens}),
one has generally the matrix element
\begin{equation}
\begin{split}
\label{eq:wesporb} \left\langle {\Psi'}^{J'M'} \right|&V \mathcal{L}
\cdot \mathcal{S}\left|
\Psi^{JM}\right\rangle = \delta_{JJ'}\delta_{MM'} (-1)^{J+S'+L} \\
\times &\begin{Bmatrix} L' & 1 & L \\ S & J & S' \end{Bmatrix}
\left\langle {\Psi'}^{K'L'} | | V \mathcal{L} | | \Psi^{KL}
\right\rangle \left\langle \chi_{S'} | | \mathcal{S} | | \chi_S
\right\rangle.
\end{split}
\end{equation}

The spin reduced element $\left\langle \chi_{S'} | | \mathcal{S} | |
\chi_S \right\rangle$ depends not only on the number $N+1$ of particles, but
also on the type of coupling; it can be calculated specifically for each
type of wave function. In any case, a closed (but somewhat complicated) formula
is obtained with successive applications of WE theorem. An example is presented
for the 3-body problem in Appendix \ref{sec:spinredel}.

In this section, we are mainly concerned with the space matrix element. The
prototype for the operator is
\begin{equation}
\label{eq:protoVLS}
V = V(| \tilde{w} \bm x |) (\tilde{w} \bm x \times \tilde{\zeta} \bm \pi),
\end{equation}
the parameters $w,\zeta$ being $w^{(i)},\zeta^{(i)}$ for one-body operators and
$w^{(ij)},\zeta^{(ij)}$ for two-body operators.

The starting point for the calculus is based on
\begin{eqnarray}
\label{eq:elmsporasint}
&\left\langle {\Psi'}^{K'L'} | | V(| \tilde{w} \bm x |) (\tilde{w} \bm x \times
\tilde{\zeta} \bm \pi) | | \Psi^{KL}\right\rangle = \int dr \, V(r)& \nonumber \\
& \left\langle {\Psi'}^{K'L'} | | \delta(| \tilde{w} \bm x |-r)  (\tilde{w}
\bm x \times \tilde{\zeta} \bm \pi) | | \Psi^{KL}\right\rangle.&
\end{eqnarray}

The reduced matrix element in the integral is calculated with the help of the
generating functions (see (\ref{eq:elmatgen})) and with the expression
(\ref{eq:elmatmomang}).

To simplify the notations, let us introduce the following notations
(we stick as close as possible to the notations adopted in SV and
SBM). First, with $B=A+A'$, one needs,
\begin{equation}
\label{eq:paramnorm}
q=\frac{1}{4} \tilde{u}B^{-1}u; \quad q'=\frac{1}{4} \tilde{u'}B^{-1}u'; \quad
\rho = \frac{1}{2} \tilde{u'}B^{-1}u.
\end{equation}
Those scalar quantities are present in the term $\mathcal{M}_0$ (see
(\ref{eq:elgennorm})) and, as such, occur in the expression for the overlap and
non relativistic kinetic energy matrix elements. Second, one must introduce
\begin{eqnarray}
\label{eq:paramvcent}
& \gamma=\frac{\tilde{w}B^{-1}u}{\tilde{w}B^{-1}w}; \quad  \gamma'=\frac
{\tilde{w}B^{-1}u'}{\tilde{w}B^{-1}w}; &\nonumber \\
& c=\frac{2}{\tilde{w}B^{-1}w}; \quad \bm z=\frac{1}{2}\tilde{w}B^{-1} \bm v.&
\end{eqnarray}
In addition to $q,q',\rho$, the scalars $c,\gamma, \gamma'$ and the vector $\bm z$
occur in the calculation of the matrix elements for central potentials. Lastly,
the quantities
\begin{equation}
\label{eq:paramvls}
\eta=\tilde{\zeta}A'B^{-1}u; \quad  \eta'=\tilde{\zeta}AB^{-1}u'
\end{equation}
are scalars specific to spin-orbit potentials.

With those definitions, it is easy to show the equalities
\begin{eqnarray}
\label{eq:egalzy}
c \bm z & = & \gamma \lambda \bm e + \gamma' \lambda' \bm e' \\
\tilde{\zeta} \bm y & = &\lambda \eta \bm e - \lambda' \eta' \bm e' \nonumber
\end{eqnarray}
so that (do not forget that $| \bm{e} | = 1 = | \bm{e}' | $)
\begin{eqnarray}
\label{eq:egal2}
c^2 z^2 & = & \gamma^2 \lambda^2 + {\gamma'}^2 {\lambda'}^2 + 2 \gamma \gamma'
\lambda \lambda' \bm e \cdot \bm e' \\
\tilde{w}B^{-1} \bm v \times \tilde{\zeta} \bm y & = & - \frac{2}{c}
(\gamma \eta' + \gamma' \eta) \lambda \lambda' \; \bm e \times \bm e'.
\nonumber
\end{eqnarray}

Let us focus on the term depending on $\lambda, \lambda', \bm e, \bm e'$ since
we must derive and integrate it. This term writes explicitly
\begin{equation}
\label{eq:termdervspor}
e^{q \lambda^2 + q' {\lambda'}^2 +\rho \lambda \lambda' \bm e \cdot
\bm e'} \; \frac{i_1(crz)}{z} e^{-c z^2/2} \lambda \lambda'
[i(\bm e \times \bm e')].
\end{equation}

At this stage one can follow two roads:
\begin{itemize}
    \item the road chosen by SV which leads to expressions in which the symmetry
properties are very simple, but less efficient numerically;
    \item the road proposed by SBM whose symmetry properties are less
transparent but more efficient numerically.
\end{itemize}

In order to give interesting and alternative expressions (very useful for
numerical checks), let us present these two roads.

In the first road, the first exponential (coming from the
$\mathcal{M}_0$ quantity) is expanded in series of its arguments,
while the $z$-depending functions are expressed in series of $z^2$
with help of formula (\ref{eq:develseril}). Lastly, the
corresponding powers of $z^2$ are expanded in powers of $\lambda,
\lambda', \bm e \cdot \bm e'$ using (\ref{eq:egal2}). All powers of
$\lambda, \lambda'$ and $\bm e \cdot \bm e'$ are then gathered. The
integration over $\bm e$ and $\bm e'$ is performed using the
expression given by (\ref{eq:prodveceep}) (forget about the
Clebsch-Gordan coefficient since it cancels  taking the reduced
matrix element). The derivation on variable $\lambda$ is easily
obtained remarking that $\partial \lambda^n /\partial \lambda^{2K+L}
|_{\lambda=0}$ = $(2K+L)! \; \delta_{2K+L,n}$ and an analogous
relation for the derivation on variable $\lambda'$. The rest of the
calculation is just matter of lengthy but straightforward algebra.
It is very interesting to note that, as it was the case for central
potential, the universal function $F^n_{p,p',l}$ (as given by
(\ref{eq:fFVarg})) also occurs in the case of spin-orbit potentials.
The final result for the reduced space matrix element writes
explicitly
\begin{equation}
\label{eq:potlsgen} \begin{split} \big\langle {\Psi'}^{K'L'} ||& V(|
\tilde{w} \bm x |) (\tilde{w} \bm x \times \tilde{\zeta} \bm \pi) |
| \Psi^{KL}\big\rangle = \delta_{L,L'}
(\gamma \eta' + \gamma' \eta)\\
\times& \sqrt{L(L+1)(2L+1)} \;
\frac{(2K+L)!(2K'+L')!}{B_{KL}B_{K'L'}}\\
\times&\left(\frac{\pi^N} {\det B}\right)^{3/2}\ \
\sum\limits_{n=0}^{K+K'+L-1} \frac{1}{c^{n+1}} J(n,c) \\
\times& \sum\limits_{k=0}^{\min(K,K')}
\frac{B_{kL}}{2k+L}F^n_{K-k,K'-k,2k+L-1} (q,q',\rho,\gamma,\gamma')
\end{split}
\end{equation}
where the dynamical integral $J(n,c)$ is given by (\ref{eq:intJnc}) with the
value of the $Q_n(x)$ function
\begin{equation}
\label{eq:Qnxls}
Q_n(x) = H_1(x) K_n^{(1)}(x)
\end{equation}
and the function $K_n^{(1)}(x)$ defined by (\ref{eq:formK}) with the special
value $l=1$.

The very important peculiar case $K=K'=0$ leads to substantial
simplifications since Eq.~(\ref{eq:potlsgen}) reduces to
\begin{eqnarray}
\label{eq:potlsK0}
&\left\langle {\Psi'}^{0L'} | | V(| \tilde{w} \bm x |) (\tilde{w} \bm x \times
\tilde{\zeta} \bm \pi) | | \Psi^{0L}\right\rangle = \delta_{L,L'} & \nonumber \\
&\sqrt{L(L+1)(2L+1)} \; \left(\frac{\eta}{\gamma} + \frac{\eta'}{\gamma'}\right)
(L-1)! \; \mathcal{N}_L& \\
&\sum\limits_{n=0}^{L-1} \frac{1}{(L-1-n)!} \; J(n,c) \left(\frac{\gamma \gamma'}
{\rho c} \right)^{n+1}& \nonumber
\end{eqnarray}

One sees that the expression is very similar to that corresponding
to central potentials (in particular the overlap $\mathcal{N}_L$,
defined in \eqref{eq:recpart}, factorizes), and this is very
interesting for numerical efficiency.

In practice, it is better to use the $\mathcal{F}$ dynamical integrals instead
of the $J$ dynamical integrals. In order to do that, the first step is to get
the series expansion of the $Q_n(x)$ function:
\begin{equation}
\label{eq:serQnls}
Q_n(x) = (-1)^n \sum\limits_{r=0}^n (-1)^r \frac{(r+1)(2x)^{2r+4}}{(2r+3)!
(n-r)!}.
\end{equation}

This expression allows to provide the link between both types of dynamical
integrals
\begin{equation}
\label{eq:lienJFls}
\begin{split}
J(n,c)&=(-1)^n \sqrt{\frac{c}{2 \pi}}\\
\times\sum\limits_{r=0}^n &(-1)^r \frac{(r+1)}{(2r+3)!(n-r)!}
(2c)^{r+2} \mathcal{F}_V(2r+4,c/2).
\end{split}
\end{equation}

Using this expression in (\ref{eq:potlsgen}), and rearranging the summations
it is possible to write the searched matrix element in term of the
$\mathcal{F}$ integrals:
\begin{equation}\begin{split}
\label{eq:potlsgen2} \big\langle & {\Psi'}^{K'L'}| | V(| \tilde{w}
\bm x |) (\tilde{w} \bm x \times
\tilde{\zeta} \bm \pi) | | \Psi^{KL}\big\rangle = \delta_{L,L'} \\
\times&\sqrt{\frac{L(L+1)(2L+1)}{\pi}} \;
\frac{(2K+L)!(2K'+L')!}{2^{K+K'}B_{KL}B_{K'L'}}
\left(\frac{2c\pi^N}{\det B}\right)^{3/2} \\
\times & \gamma^{2K+L} {\gamma'}^{2K'+L'} \left(\frac{\eta}{\gamma}
+ \frac{\eta'}{\gamma'}\right) \left( - \frac{1}{c}
\right)^{K+K'+L-1} \\
\times&\sum\limits_{n=0}^{K+K'+L-1} \frac{(n+1)}{(2n+3)!} (-2c)^n
\mathcal{F}_V(2n+4,c/2) \\
\times & \sum\limits_{k=0}^{\min(K,K')} 4^k \frac{B_{k\,L}}{2k+L}
H_{n,k}^{K,K',L-1} \left( \frac{2q \gamma'}{\rho \gamma},\frac{2q'
\gamma}{\rho \gamma'}, \frac{\rho c}{\gamma \gamma'} \right).
\end{split}
\end{equation}

Here again, it is very pleasant to ascertain that it appears in this expression
the same geometrical $H$ function (\ref{eq:fHnk}) as in the central potential
expression (compare to (\ref{eq:Vcentgen2})).

For the peculiar important case $K=K'=0$, the matrix element has a much simpler
form:
\begin{equation}\begin{split}
\label{eq:potls2K0} \big\langle & {\Psi'}^{0L'} | | V(| \tilde{w}
\bm x |) (\tilde{w} \bm x \times \tilde{\zeta} \bm \pi) | |
\Psi^{0L}\big\rangle = \delta_{L,L'}  \; (L-1)!  \;
\mathcal{N}_L\\
\times&\sqrt{\frac{L(L+1)(2L+1)}{8 \pi c}}\left(\frac{\eta}{\gamma}
+ \frac{\eta'}{\gamma'}\right)
 \sum\limits_{n=0}^{L-1} \mathcal{F}_V(2n+4,c/2)   \\
\times & \frac{(n+1)\;(2c)^{n+3}} {(2n+3)!(L-1-n)!} \left(\frac{
\gamma \gamma'}{\rho c} \right)^{n+1} \left(1 - \frac{\gamma
\gamma'} {\rho c}\right)^{L-1-n}.
\end{split}
\end{equation}

This expression is very similar to (\ref{eq:potlsK0}) (at least as simple) but
has the big advantage to be given in terms of $\mathcal{F}$ integrals instead
of the more complicated $J$ integrals; thus this expression is much more suited
for a numerical code.

In the second road, we take opportunity of the link between $\bm e \cdot \bm e'$
and $z^2$ given in eq. (\ref{eq:egal2}) to transform the fist exponential in
(\ref{eq:termdervspor}) under a form containing $z^2$ instead of $\bm e \cdot
\bm e'$. The part depending on $z^2$ is then gathered with the second
exponential in (\ref{eq:termdervspor}). This trick allows to gain one series
expansion in the development of (\ref{eq:termdervspor}) and, thus, a summation
index less than in the former treatment. Indeed, instead of
(\ref{eq:termdervspor}), our expansion is based on the alternative form
\begin{equation}
\label{eq:termdervspor2}
e^{\bar{q} \lambda^2 + \bar{q'} {\lambda'}^2}  \; \frac{i_1(crz)}{z}
e^{- \alpha c z^2/2} \lambda \lambda' \; [i(\bm e \times \bm e')].
\end{equation}
with introduction of new parameters
\begin{equation}
\label{eq:quantbar}
\bar{q} = q-\frac{\rho \gamma}{2 \gamma'}; \quad \bar{q}\,' = q'-
\frac{\rho \gamma'}{2 \gamma}; \quad \alpha = 1 - \frac{\rho c}{\gamma \gamma'}.
\end{equation}

The rest of the derivation is quite similar to what was done previously.

Thus we have a new expression for the matrix element
\begin{equation}\begin{split}
\label{eq:potlsgen3} \big\langle &{\Psi'}^{K'L'} | | V(| \tilde{w}
\bm x |) (\tilde{w} \bm x \times \tilde{\zeta} \bm \pi) | |
\Psi^{KL}\big\rangle = \delta_{L,L'}
(\gamma \eta' + \gamma' \eta) \\
\times&\sqrt{L(L+1)(2L+1)} \;
\frac{(2K+L)!(2K'+L')!}{B_{KL}B_{K'L'}}\left(
\frac{\alpha \pi^N}{\det B}\right)^{3/2} \\
\times&\sum\limits_{n=0}^{K+K'+L-1}
\left(\frac{\alpha}{2c}\right)^{n+1}
J(n,\alpha,c) \\
\times& \sum\limits_{k=0}^{\min(K,K')}
2^{2k+L}\frac{B_{k\,L}}{(2k+L)!}
F^{K,K',L-1}_{n,k}(\bar{q},\bar{q'},\gamma,\gamma').
\end{split}
\end{equation}
where the integral $J(n,\alpha,c)$ is defined by (\ref{eq:intJnac}) with the same
function $Q_n(x)$ as before (\ref{eq:Qnxls}).

Once more, the universal function $F$, given by (\ref{eq:fFmoi}), appears
naturally. As explained, this expression is better suited for numerical
calculation as compared to the analoguous formulation (\ref{eq:potlsgen}).

The peculiar case $K=K'=0$ is particularly simple
\begin{equation}\begin{split}
\label{eq:potls3K0} \big\langle {\Psi'}^{0L'} &| | V(| \tilde{w} \bm
x |) (\tilde{w} \bm x \times
\tilde{\zeta} \bm \pi) | | \Psi^{0L}\big\rangle = \delta_{L,L'} \\
\times &\sqrt{L(L+1)(2L+1)} \;
\left(\frac{\eta}{\gamma}+\frac{\eta'}{\gamma'}\right)
(L-1)! \; \mathcal{N}_L \\
\times&\alpha^{3/2} \left(\frac{\alpha}{1-\alpha}\right)^L
J(L-1,\alpha,c). \end{split}
\end{equation}

Since the integral $J(L-1,\alpha,c)$ needs essentially the same numerical effort
than the integral $J(n,c)$, it is obvious that the new formula
(\ref{eq:potls3K0}) is much efficient than its analog (\ref{eq:potlsK0}); it is
a closed analytical expression without any summation!

Let us close this section by giving new expressions in terms of the dynamical
$\mathcal{F}$ integrals. Always with the help of (\ref{eq:serQnls}), the link
between the $J$ integrals and the $\mathcal{F}$ integrals is given by
\begin{equation}\begin{split}
\label{eq:lienJFls2}
J(n,\alpha,c)&=(-1)^n \sqrt{\frac{c}{2 \pi \alpha}} \\
\times\sum\limits_{r=0}^n  &\frac{(-1)^r(r+1)}{(2r+3)!(n-r)!}\left(
\frac{2c} {\alpha}\right)^{r+2} \mathcal{F}_V(2r+4,c/2).
\end{split}
\end{equation}

The final expression for the general matrix element looks similar to
\begin{equation}\begin{split}
\label{eq:potlsgen3} \big\langle &{\Psi'}^{K'L'} | | V(| \tilde{w}
\bm x |) (\tilde{w} \bm x \times \tilde{\zeta} \bm \pi) | |
\Psi^{KL}\big\rangle = \delta_{L,L'} (\gamma \eta'+
\gamma' \eta) \\
\times&\sqrt{\frac{L(L+1)(2L+1)}{4 \pi}} \;
\frac{(2K+L)!(2K'+L')!}{B_{KL}B_{K'L'}}
\left(\frac{2c\pi^N}{\det B}\right)^{3/2} \\
\times & \sum\limits_{n=0}^{K+K'+L-1} \frac{(n+1)}{(2n+3)!} \mathcal{F}_V(2n+4,c/2) \\
\times & \sum\limits_{k=0}^{\min(K,K')} 2^{2k+L}
\frac{B_{k\,L}}{(2k+L)!} P_{n,k}^{K,K',L-1} \left( \bar{q
},\bar{q'},\gamma,\gamma',\frac{- \alpha {\gamma'}^2}{2 c \bar{q'}}
\right).\end{split}
\end{equation}

The $P$ function has been defined in (\ref{eq:fPnk}).

Application of this formula to the special case $K=K'=0$ does not bring
anything new since it reduces to the one already got previously (see
(\ref{eq:potls2K0}))

\section{Tensor force}
\label{sec:tensfor}
In few and many-body systems, the tensor force is of common use. We discuss
here the two-body tensor force $V^{(T)}=\sum_{i \leq j} V^{(T)}_{ij}
(\bm r_{ij})$.

The form of this potential for the $(i-j)$ pair is traditional
\begin{equation}
\label{eq:formtens} V^{(T)}_{ij}(\bm r_{ij})=V(| \bm r_{ij} |)
\widehat{S}_{ij}; \quad \widehat{S}_{ij} = 3 (\bm s_i \cdot
\widehat{\bm r}_{ij})(\bm s_j \cdot \widehat{\bm r}_{ij}) - \bm s_i
\cdot \bm s_j.
\end{equation}

Very often we find an alternative form for the tensor operator namely
\begin{equation}
\label{eq:formtens2} \widehat{S}_{ij} = 3 (\bm S_{ij} \cdot
\widehat{\bm r}_{ij})^2 - \bm S_{ij}^2.
\end{equation}
where the total spin of the pair $\bm S_{ij}=\bm{s}_i + \bm{s}_j$
enters the game. Strictly speaking both forms are fully equivalent
(up to a factor 2) only for spin 1/2 particles. However, in deriving
the interaction between particles it appears, in addition to the
form (\ref{eq:formtens}), terms depending on $(\bm{s}_i \cdot
\widehat{\bm r}_{ij})^2$ and $(\bm{s}_j \cdot \widehat{\bm
r}_{ij})^2$ which can be absorbed in the former expression to give a
term similar to (\ref{eq:formtens2}) (up to uninteresting
constants). A more detailed discussion about these forms for the
tensor force can be found in ref. \cite{Mathieu:2007mw}. Thus, both
forms can be considered as equivalent and are of common use.

To treat easily the tensor force, it is better to express it in a form that
splits the space and spin degrees of freedom. It is well known that this operator
can be recast under form (the dot means the scalar product, while the bracket
means the coupling to 0 angular momentum).
\begin{eqnarray}
\label{eq:opSij}
\widehat{S}_{ij} & = &\sqrt{\frac{24 \pi}{5}}
\left(Y_2(\widehat{\bm r}_{ij}) \cdot \mathcal{S}_{ij}
\right) \nonumber \\
 & = & \sqrt{24 \pi} \left[Y_2(\widehat{\bm r}_{ij}) \otimes
\mathcal{S}_{ij} \right]_{00}
\end{eqnarray}
with one of the alternative expression
\begin{equation}
\label{eq:altformSij}
 \mathcal{S}_{ij} = (\bm s_i \otimes \bm s_j)_2 \quad \textrm{or} \quad
\mathcal{S}_{ij}=(\bm S_{ij} \otimes \bm S_{ij})_2.
\end{equation}

Using again (\ref{eq:elmatprodtens}), one has
\begin{equation}\begin{split}
\label{eq:wetens} \left\langle {\Psi'}^{J'M'}
\right|&V^{(T)}_{ij}\left|
\Psi^{JM} \right\rangle = \delta_{JJ'}\delta_{MM'} (-1)^{J+S'+L}  \\
\times & \sqrt{\frac{24 \pi}{5}} \begin{Bmatrix} L' & 2 & L \\ S & J
& S' \end{Bmatrix}\left\langle \chi_{S'} | | \mathcal{S}_{ij}
| | \chi_S \right\rangle
\\ \times & \left\langle {\Psi'}^{K'L'} | | V(| \bm r_{ij} |) Y_2(\widehat{\bm
r_{ij}})| | \Psi^{KL} \right\rangle .
\end{split}
\end{equation}

The spin matrix elements $\left\langle \chi_{S'} | | \mathcal{S}_{ij}
| | \chi_S \right\rangle$ must be calculated separately for each type of
coupling for a given system. Anyhow, this calculation can always be performed
by successive application of Wigner-Eckart theorem and with the help of
(\ref{eq:elmatprodtens}). The corresponding formulae for the three-body
systems are presented in Appendix \ref{sec:spinredel} (see
(\ref{eq:redmatspintens1}-\ref{eq:redmatspintens4})).

In the rest of this section, we focus on the space matrix element. Because of
(\ref{eq:comblinpos}), we are concerned with the matrix element
\begin{equation}\begin{split}
\label{eq:elmsporasint} \big\langle &{\Psi'}^{K'L'} | | V(|
\tilde{w} \bm x |) Y_2(\widehat{\tilde{w}
\bm x}) | | \Psi^{KL}\big\rangle = \\
& \int dr \, V(r)\left\langle {\Psi'}^{K'L'} | | \delta(| \tilde{w}
\bm x |-r) Y_2(\widehat{\tilde{w}\bm x})  | |
\Psi^{KL}\right\rangle.\end{split}
 \end{equation}

The derivation is quite similar to the one presented in the case of
spin-orbit potential and we do not repeat the arguments. Let us just
point out a trick that was used; the term $Y_{2 \mu}(\widehat{2 \bm
z)}$ that appears can be transformed to $Y_{2 \mu}(\hat{\bm z)}$  =
$\mathcal{Y}_{2 \mu}(c \bm z) /(c^2 z^2)$. With the expression
(\ref{eq:egalzy}) for $c \bm z$, and the values (\ref{eq:reslamb2})
for the integral on the variables $\bm e, \bm e'$, the rest of the
calculation is straightforward.

Let us give first the matrix element obtained by ``the first road''. The most
general expression is
\begin{equation}\begin{split}
\label{eq:pottensgen} \big\langle &{\Psi'}^{K'L'} | | V(| \tilde{w}
\bm x |) Y_2(\widehat{\tilde{w}
 \bm x}) | | \Psi^{KL}\big\rangle = \delta_{L \pm 2 \textrm{ or }L,L'} \\
\times&\sqrt{\frac{5}{4 \pi}} \;
\frac{(2K+L)!(2K'+L')!}{B_{KL}B_{K'L'}} \left(\frac{\pi^N} {\det
B}\right)^{3/2} \gamma \gamma'
 \\
\times & \sqrt{\frac{L_m (L_m+1)}{2L_m
+1}}\sum\limits_{n=0}^{K+K'+L_m-1} \frac{1}{c^{n+1}} J(n,c)
 \\
\times&  \sum\limits_{l=0}^{2}
\sum\limits_{k=0}^{\min(\bar{K}_l,\bar{K'}_l)}
\mathcal{A}^{L,L'}_{kl}(\gamma,\gamma')
F^n_{\bar{K}_l-k,\bar{K'}_l-k,2k+\bar{L}_l}
(q,q',\rho,\gamma,\gamma') \end{split}
\end{equation}
where the dynamical integral $J(n,c)$ is given by (\ref{eq:intJnc}) with the
value of the $Q_n(x)$ function
\begin{equation}
\label{eq:Qnxtens}
Q_n(x) = \frac{K_n^{(2)}(x)}{x}
\end{equation}
and the function $K_n^{(l)}(x)$ defined by (\ref{eq:formK}) with the
special value $l=2$. The geometrical coefficients $\mathcal{A}$ are
given by
\begin{subequations}\label{eq:coefAkl}
\begin{eqnarray}
\mathcal{A}^{L,L \pm 2}_{kl}(\gamma,\gamma') & = &\sqrt{\frac{3}{2}}
2^{\bar{l}}
\left(R_{\pm} \right)^{l-1}\; B_{k\,(L_i+l)}; \\
\mathcal{A}^{L,L}_{kl}(\gamma,\gamma') & = &- \frac{2L+1}{\sqrt{(2L-1)(2L+3)}}
\left(R_+ \right)^{l-1} \nonumber \\
 & & \times B_{k\,L} \frac{(4k+2L+2+\bar{l})}{2^{1 - \bar{l}}(2k+L+1-\bar{l})};
\\
R_+ =\gamma/\gamma' & ; & R_- = \gamma'/\gamma; \nonumber
\end{eqnarray}
\end{subequations}
where new quantities are introduced below
\begin{eqnarray}
\label{eq:defKbar}
\bar{K}_l=K-l; \; \bar{K'}_l=K'; \; \bar{L}_l = L+l & \textrm{ for }& L'=L+2
\nonumber \\
\bar{K}_l=K-f_l; \; \bar{K'}_l=K'-f'_l; \; \bar{L}_l = L - \bar{l} &
\textrm{ for }& L'=L \\
\bar{K}_l=K; \; \bar{K'}_l=K'-l; \; \bar{L}_l = L-2+l & \textrm{ for }& L'=L-2
\nonumber
\end{eqnarray}
and
\begin{equation}
\label{eq:Lmi}
L_m = (L+L')/2; \; L_i = \min(L,L'); \; \bar{l}= \textrm{mod}(l,2);
\end{equation}
\begin{equation}
\label{eq:flfpl}
f_l = (l - \bar{l})/2; \quad f'_l = (2 - l - \bar{l})/2.
\end{equation}

For the special case $K=K'=0$ the corresponding expressions are much simpler
\begin{equation}\begin{split}
\label{eq:pottensK0} \big\langle &{\Psi'}^{0L'} | | V(| \tilde{w}
\bm x |) Y_2(\widehat{\tilde{w}
 \bm x}) | | \Psi^{0L}\big\rangle = \delta_{L \pm 2 \textrm{ or }L,L'} \\
\times &\sqrt{\frac{5}{4 \pi}} \; \sqrt{L_m(L_m+1)} \;
\mathcal{N}_{L_m} \; (L_m-1)!
\; \mathcal{D}^{L,L'}(\gamma,\gamma') \\
\times & \sum\limits_{n=0}^{L_m-1} \frac{1}{(L_m-1-n)!} J(n,c)
\left( \frac{\gamma \gamma'}{\rho c} \right)^{n+1}\end{split}
\end{equation}
and
\begin{subequations}\label{eq:defcoefD}
\begin{eqnarray}
 \mathcal{D}^{L,L \pm
2}(\gamma,\gamma')&=&(2L_m+3) \sqrt{\frac{3}{2(2L_m+1)}} R_{\mp};
 \\
\mathcal{D}^{L,L}(\gamma,\gamma')&=& -
\sqrt{\frac{(2L+1)(2L+3)}{(2L-1)}}.
\end{eqnarray}
\end{subequations}

The alternative expression in terms of the $\mathcal{F}$ integrals is directly
obtained using the link between the $J$ and $\mathcal{F}$ integrals:
\begin{equation}\begin{split}
\label{eq:lienJFtens}
J(n,c)= \frac{(-1)^n}{\sqrt{2 \pi c}}& \sum\limits_{r=0}^n (-1)^r \frac{(r+1)(r+2)}{(2r+5)!(n-r)!}\\
\times& (2c)^{r+3} \mathcal{F}_V(2r+4,c/2).\end{split}
\end{equation}

The final result is
\begin{equation}\begin{split}
\label{eq:pottensgen2} \big\langle& {\Psi'}^{K'L'} | | V(| \tilde{w}
\bm x |) Y_2(\widehat{\tilde{w}
 \bm x}) | | \Psi^{KL}\big\rangle = \delta_{L \pm 2 \textrm{ or }L,L'}\\
\times&\sqrt{\frac{5}{\pi}} \; \sqrt{\frac{L_m (L_m+1)}{\pi (2L_m
+1)}} \; \frac{(2K+L)!(2K'+L')!}{2^{K+K'} B_{KL}B_{K'L'}}
\\
\times&\left(\frac{2 c \pi^N} {\det B}\right)^{3/2} \gamma^{2K+L}
{\gamma'}^{2K'+L'} \left(- \frac{1}{c} \right)^{K+K'+L_m - 1} \\
\times&\sum\limits_{n=0}^{K+K'+L_m-1} \frac{(n+1)(n+2)}{(2n+5)!}
(-2c)^n
\mathcal{F}_V(2n+4,c/2)  \\
\times &\sum\limits_{l=0}^{2}
\sum\limits_{k=0}^{\min(\bar{K}_l,\bar{K'}_l)}
 4^k \mathcal{B}^{L,L'}_{k\,l} H^{\bar{K}_l,\bar{K'}_l,\bar{L}_l}_{n,k}
\left( \frac{2q \gamma'}{\rho \gamma},\frac{2q' \gamma}{\rho
\gamma'}, \frac{\rho c}{\gamma \gamma'} \right). \end{split}
\end{equation}

The geometrical coefficients $\mathcal{B}^{L,L'}_{kl}$ are independent of
any parameter and looks similar to
\begin{eqnarray}
\label{eq:coefBkl} \mathcal{B}^{L,L \pm 2}_{k\,l} & =
&\sqrt{\frac{3}{2}} 2^{l+\bar{l}} \;
B_{k\,(L_i+l)}; \\
\mathcal{B}^{L,L}_{k\,l} & = &- \frac{2L+1}{\sqrt{(2L-1)(2L+3)}} \;
B_{k\,L} \frac{(4k+2L+2+\bar{l})}{(2k+L+1-\bar{l})}; \nonumber
\end{eqnarray}

One sees the great similarity between (\ref{eq:pottensgen2}) and
(\ref{eq:potlsgen2}), showing once more the universality of the
$H$-function.

The expression for the peculiar case $K=K'=0$ is easily derived from
(\ref{eq:pottensK0}) and (\ref{eq:lienJFtens}) after some algebraic
manipulations
\begin{equation}\begin{split} \label{eq:pottensK02} \big\langle& {\Psi'}^{0L'} | | V(|
\tilde{w} \bm x |) Y_2(\widehat{\tilde{w}
 \bm x}) | | \Psi^{0L}\big\rangle = \delta_{L \pm 2 \textrm{ or }L,L'} \\
\times &\sqrt{\frac{5}{ \pi}} \; \sqrt{\frac{L_m(L_m+1)}{8 \pi c}}
\; \mathcal{N}_{L_m}
\; (L_m-1)! \; \mathcal{D}^{L,L'}(\gamma,\gamma') \\
\times& \sum\limits_{n=0}^{L_m-1} \frac{(n+1)(n+2)}{(2n+5)!
(L_m-1-n)!}
(2c)^{n+3} \mathcal{F}_V(2n+4,c/2) \\
 \times& \left( \frac{\gamma \gamma'}{\rho c} \right)^{n+1}
\left(1 - \frac{\gamma \gamma'}{\rho c}\right)^{L_m -
(n+1)}\end{split}
\end{equation}
with the same geometrical coefficients $\mathcal{D}^{L,L'}$ as before
(\ref{eq:defcoefD}).

The alternative derivation can be obtained using exactly the same technics that
was developed for the spin-orbit potential. The explicit result looks quite
sympathetic:
\begin{equation}\begin{split}
\label{eq:pottensgen3} \big\langle& {\Psi'}^{K'L'} | | V(| \tilde{w}
\bm x |) Y_2(\widehat{\tilde{w}
 \bm x}) | | \Psi^{KL}\big\rangle = \delta_{L \pm 2 \textrm{ or }L,L'}\sqrt{\frac{5}{\pi}} \gamma \gamma'\\
\times& \; \frac{(2K+L)!(2K'+L')!}{B_{KL}B_{K'L'}} \left(
\frac{\alpha \pi^N} {\det B}\right)^{3/2} \sqrt{\frac{L_m
(L_m+1)}{2L_m
+1}}\\
\times &\sum\limits_{n=0}^{K+K'+L_m-1}
\left(\frac{\alpha}{2c}\right)^{n+1} J(n,\alpha,c)
\sum\limits_{l=0}^{2} \sum\limits_{k=0}^{\min(\bar{K}_l,\bar{K'}_l)} \\
\times &  \frac{2^{2k+\bar{L}_l}}{(2k+\bar{L}_l)!
}\mathcal{A}^{L,L'}_{kl} (\gamma,\gamma')
F^{\bar{K}_l,\bar{K'}_l,\bar{L}_l}_{n,k} (\bar{q},\bar{q'},
\gamma,\gamma')\end{split}
\end{equation}
where the dynamical integral $J(n,\alpha,c)$ is given by
(\ref{eq:intJnac}) with the same value of the $Q_n(x)$ as before
(\ref{eq:Qnxtens}). The coefficients $\mathcal{A}^{L,L'}_{kl}$ are
still given by (\ref{eq:coefAkl}) and the indices
$\bar{K}_l,\bar{K'}_l,\bar{L}_l$ by (\ref{eq:defKbar}). The
$F$-function (\ref{eq:fFmoi}) appears naturally in this expression.

The special case $K=K'=0$ is particularly simple since it does not need any
summation
\begin{equation}\begin{split}
\label{eq:pottensK03} \big\langle &{\Psi'}^{0L'} | | V(| \tilde{w}
\bm x |) Y_2(\widehat{\tilde{w}
 \bm x}) | | \Psi^{0L}\big\rangle = \delta_{L \pm 2 \textrm{ or }L,L'} \\
\times&\sqrt{\frac{5}{4 \pi}} \; \sqrt{L_m(L_m+1)} \;
\mathcal{N}_{L_m} \; (L_m-1)! \\
\times & \mathcal{D}^{L,L'}(\gamma,\gamma') \alpha^{3/2} \left(
\frac{\alpha}{1 - \alpha} \right)^{L_m} J(L_m-1,\alpha,c)\end{split}
\end{equation}
with the same geometrical coefficients $\mathcal{D}^{L,L'}$ as before
(\ref{eq:defcoefD}).

The last thing that remains to do is to express the reduced matrix elements
in terms of the $\mathcal{F}$ integrals. This is easily done using the
expansion
\begin{equation}\begin{split}
\label{eq:lienJFtens2} J(n,\alpha,c)=& (-1)^n\sqrt\frac{\alpha}{2
\pi c} \sum\limits_{r=0}^n \mathcal{F}_V(2r+4,c/2)\\
&\times  (-1)^r \frac{(r+1)(r+2)}{(2r+5)!(n-r)!} \left(\frac{2c}
{\alpha}\right)^{r+3}.\end{split}
\end{equation}

One gets
\begin{equation}\begin{split}
\label{eq:pottensgen4} \big\langle &{\Psi'}^{K'L'} | | V(| \tilde{w}
\bm x |) Y_2(\widehat{\tilde{w}
 \bm x}) | | \Psi^{KL}\big\rangle = \delta_{L \pm 2 \textrm{ or }L,L'}
\sqrt{\frac{5}{\pi}} \\
\times &\sqrt{\frac{L_m (L_m+1)}{\pi (2L_m +1)}} \;
\frac{(2K+L)!(2K'+L')!}{B_{KL}B_{K'L'}} \left(\frac{2 c \pi^N}
{\det B}\right)^{3/2} \gamma \gamma' \\
\times &\sum\limits_{n=0}^{K+K'+L_m-1} \frac{(n+1)(n+2)}{(2n+5)!}
\mathcal{F}_V(2n+4,c/2)  \sum\limits_{l=0}^{2} \sum\limits_{k=0}
^{\min(\bar{K}_l,\bar{K'}_l)} \\ \times &
 \frac{2^{2k+\bar{L}_l}}{(2k+\bar{L}_l)!} \mathcal{A}^{L,L'}_{kl}(\gamma,
 \gamma') P^{\bar{K}_l,\bar{K'}_l,\bar{L}_l}_{n,k}
\left(\bar{q},\bar{q'},\gamma,\gamma',\frac{- \alpha {\gamma
'}^2}{2c\bar{q'}} \right)\end{split}
\end{equation}
where, again, the $P$ function (\ref{eq:fPnk}) enters the game.

Starting with (\ref{eq:pottensK03}), using (\ref{eq:lienJFtens2})
and rearranging the summations allows to obtain the matrix element
for the special case $K=K'=0$ under the form already proposed
(\ref{eq:pottensK02}). This is a fantastic check of the calculation.

\section{Conclusions}
\label{sec:concl}
In this paper, we pursued the work begun in SV and SBM and proposed general
expressions for the matrix elements of spin dependent operators on correlated
gaussian wave functions with natural parity. This type of basis wave functions
are the basic ingredients of the stochastic variational method which allows
to get very precise results for few-body systems.

The corresponding formulae are able to treat the matrix elements of
the Hamiltonian for a system with an arbitrary number of particles
and for states with arbitrary angular momentum in a closed form.
This is very important from the numerical point of view; the only
purely numerical work is the computation of a one dimensional
integral containing the form of the potential. Moreover, for most of
the usual potentials, the corresponding integrals are themselves
analytical, so that all the matrix elements are obtained in a closed
form. The solution for the Schr\"{o}dinger equation is subsequently
obtained as a generalized eigenvalue problem which is very well
under control numerically.

The only spin dependent potentials that are treated in this paper are the
spin-orbit -- symmetric and antisymmetric-- and tensor forces. They are, by far,
the most common for atomic, nuclear and hadronic spectroscopy. The spin
matrix elements are computed with standard Racah algebra and we focus here on
space reduced matrix elements.

We proposed two ways to calculate the matrix elements~: one approach
based on the underlying philosophy of SV where the symmetry
properties are obvious, and another approach developed in SBM where
the symmetry properties are less transparent but more efficient
numerically. Since the link between both is far from obvious, a
comparison between both formulations is indeed a very good check of
the numerical codes. We gave the formulae for the general correlated
Gaussians $K \neq 0$ but also for the special case $K = 0$ for which
they are much simpler. In this case, all the matrix elements have
the remarkable feature to be proportional to the overlap matrix
element.

In each approach, we also proposed two expressions
\begin{itemize}
    \item one based on a numerical integral of type $J$ which lead to the simplest
formulation but which has the drawback to be given in terms of combinations of
Hermite polynomials;
    \item one based on a numerical integral of type $\mathcal{F}$ which gives a
slightly more complicated formulation but with an easy and universal type of
integral.
\end{itemize}
In the special case $K = 0$, both formulations are of the same difficulty so
that the second approach is much more convenient.

Again, all these alternative expressions for the computation of the matrix
elements allow very good checks of the results.

The very sympathetic feature of the method proposed in this paper is that
it relies on universal geometrical functions that can be built once and for all
and that can be employed whatever the potential under consideration. Moreover
the arguments that enter these functions depend on the free parameters of the
basis wave function and the particular pair of particles under consideration
only and are independent of the form of the potential for a given pair.
Consequently all types or combinations of potentials can be treated on the
same footing in a single type of summation. This is very important to shorten
the computer time needed for the evaluation of the matrix elements.

\smallskip

\acknowledgments \label{ackno} We are indebted to C. Semay for a
careful reading of our manuscript, and we are very grateful to Prof.
Y. Suzuki for fruitful remarks and advices. One of the authors
(V.M.) thanks IISN for financial support.

\appendix
\section{some matrix elements for generating functions}
In order to calculate matrix elements of some operator in the basis of
correlated Gaussians, the first step is to calculate the matrix elements of
this operator on the generating functions. Expressing the coordinates of the
particles in terms of the Jacobi coordinates $\bm x$, the most general form
for the spatial part of the operator is $V(\tilde{w}\bm x)$. Thus, it
is natural to compute the matrix element of the operator $\delta(\tilde{w}
\bm x - \bm r)$. Such an expression can be found in SV.

However, it appears that, in any case, the form of the operator is
rather $V(| \tilde{w}\bm x|) \times F(\tilde{w}\bm x)$ where the
function $F(\tilde{w}\bm x)$ is very specific and given once for
all. The element can be calculated from the expression in term of
three dimensional Dirac function, but this needs evaluating a three
dimensional integral. We find more convenient to calculate the
matrix elements on generating functions for an operator of type
$\delta(| \tilde{w}\bm x|-r) \times F(\tilde{w}\bm x)$. The angular
integration is reported entirely on the specific function $F$ and
the remaining job is just a one dimensional radial integral. The
resulting expressions are not given in SV, and we think that they
can be interesting for the reader.

\subsection{Case of spherical harmonics}
\label{subsec:spherharm} A common case concerns spherical harmonics
$F(\tilde{w}\bm x) = Y_{\lambda\mu}(\widehat{\tilde{w}\bm x})$. The
central potential corresponds to the case $\lambda=0$, while the
tensor case corresponds to $\lambda=2$.

The technics to calculate the matrix element is based on a
well-known trick. The generating functions are expressed in terms of
Gaussians and grouped into a single Gaussian of the form
$\exp(-\tilde{\bm x} \cdot B \bm x + \tilde{\bm v} \cdot \bm x)$,
where $B = A + A'$ and $\bm v = \bm s + \bm s'$. The $B$ matrix,
which is symmetric and definite positive, is diagonalized to a
matrix $D$ with help of an orthogonal matrix $T$. Instead of $\bm x$
variables, we use new variables $\bm z$, defined by $\bm
z=D^{1/2}\tilde{T} \bm x$. The exponential takes the form
$\exp(-\tilde{\bm z} \cdot \bm z + \tilde{\bm u} \cdot \bm z)$ and
the argument of the Dirac function becomes $| \tilde{a} \bm z | -
r$. We then change again variables to $\bm Z = U \bm z$, where $U$
is an orthogonal matrix so that $\tilde{\bm z} \cdot \bm z =
\tilde{\bm Z} \cdot \bm Z$. One can use the freedom left to the form
of $U$ to choose a peculiar form such as $\bm Z_1$ is proportional
to $\tilde{a} \bm z$ (the interested reader can refer to BSM for the
notations). The rest of the derivation is standard and
straightforward. To perform the angular integration for the
spherical harmonic, it is convenient to expand the corresponding
$\exp(\bm V_1 \cdot \bm Z_1)$ term as the usual plane wave
development in terms of spherical harmonics; this one introduces the
spherical modified Bessel function $i_l(z) = \sqrt{\pi/(2z)}
I_{l+1/2}(z)$.

The final result is
\begin{equation}\begin{split}
\label{eq:elmatdiryl} \left\langle g(\bm s', A'; \bm x) | \delta(|
\tilde{w}\bm x|-r) Y_{\lambda\mu}\left(\widehat{\tilde{w}\bm
x}\right) | g(\bm s, A; \bm x)
\right\rangle = \\
\frac{4}{\sqrt{\pi}} \frac{\mathcal{M}_0}{(\tilde{w} B^{-1}
w)^{3/2}} r^2 \mathcal{I}_\lambda(w,B,\bm v)
Y_{\lambda\mu}\left(\widehat{\tilde{w}B^{-1} \bm
v}\right),\end{split}
\end{equation}
in which
\begin{equation}
\label{eq:elgennorm}
\mathcal{M}_0=\left(\frac{\pi^N}{\det B}\right)^{3/2} \exp\left(\frac{1}{4}
\tilde{\bm v \cdot B^{-1} \bm v}\right)
\end{equation}
is the matrix element corresponding to the overlap of generating functions
with
\begin{equation}
\label{eq:defBv}
B = A + A' \quad ; \quad \bm v = \bm s + \bm s'
\end{equation}
and
\begin{equation}
\mathcal{I}_\lambda(w,B,\bm v)=i_\lambda \left(\frac{r |
\tilde{w}B^{-1} \bm v |}{\tilde{w} B^{-1} w}\right) \exp \left(-
\frac{r^2 +\frac{1}{4} (\tilde{w}B^{-1}\bm v)^2}{\tilde{w} B^{-1}
w}\right).
\end{equation}

\subsection{Angular momentum}
\label{subsec:angmom}
In the case of angular momentum, the $F$ function is a bit more complicated
since it contains, in addition to the Jacobi variables, the derivatives of
them. Explicitly, it is the vector product $F=\tilde{w} \bm x \times
\tilde{\zeta} \bm \pi$. The momentum $\bm \pi = -i \partial/\partial \bm x$
needs to calculate the derivative of the generating ket function. Fortunately
this derivative is still proportional to a Gaussian, so that the same
procedure as before can be adopted. There are additional terms which can
be treated exactly. The vector product being a tensor of order 1, it is
expected that the $i_1$ function appears; this is indeed the case.

Explicitly, one finds
\begin{equation}\begin{split}
\label{eq:elmatmomang} \left\langle g(\bm s', A'; \bm x) | \delta(|
\tilde{w}\bm x|-r) ( \tilde{w} \bm x \times \tilde{\zeta} \bm \pi) |
g(\bm s, A; \bm x)
\right\rangle = -i \frac{4}{\sqrt{\pi}} \\
\times\frac{\mathcal{M}_0\;r^3}{(\tilde{w} B^{-1} w)^{3/2}|
\tilde{w}B^{-1}\bm v |}  \mathcal{I}_1(w,B,\bm v)
[(\tilde{w}B^{-1}\bm v) \times (\tilde{\zeta} \bm y)], \end{split}
\end{equation}
where, in addition to the quantities previously defined, we have a new
variable
\begin{equation}
\label{eq:defy}
\bm y = A'B^{-1} \bm s - A B^{-1} \bm s'.
\end{equation}

One sees that in both cases, we have more or less the same numerical quantities
to compute, and this is a very important feature in the efficiency of the
numerical codes.

\section{A special series expansion}
\label{sec:specexp} During the calculation of the interesting matrix
elements, it is very helpful to have a series expansion for the
function $\exp(-a z^2) i_l(bz)$. The case $l=0$ corresponds to a
central potential, while $l=1$ appears for a spin-orbit potential.
The expression  for $l=0$ is given in SV (formula (A.125)) while
expression for $l=1$ is also in SV (formula (A.164)).

However, for the tensor operator, we need the expression for $l=2$. This
expression is missing in SV. Moreover the expressions proposed in SV are
under a form that is not transparent for generalization. To fill this gap,
we propose here an interesting formula, valid for any $l$. It is expressed in
terms of Hermite polynomials of odd order $H_{2n+1}$. These polynomials occur
naturally since they come as by-product of their generating function.
More precisely
\begin{equation}
\label{eq:genhermit}
e^{-s^2+2sx}=\sum_{n=0}^\infty H_n(x) \frac{s^n}{n!}.
\end{equation}

The first thing to do is to use a new variable $u =2 \sqrt{a} z$, so that
$\exp(-a z^2) i_l(bz) = \exp(-u^2/4) i_l(\alpha u)$ is now function of only
one parameter $\alpha = b/(2\sqrt{a})$. The series expansion of $i_l(x)$ is
a sum of a polynomial in $x$ times a term like $e^x$ and another polynomial
in $x$ times a term like $e^{-x}$, both divided by some power of $x$
(see ref. \cite{abra65}). For small values of $l$ these polynomials are not
complicated and formula (\ref{eq:genhermit}) can be applied safely. One can
recast the searched expression under the form:

\begin{equation}
\label{eq:expdevelopil}
\exp(-u^2/4) i_l(\alpha u) = \frac{1}{2\alpha^{l+1}} \sum_{p=0}^\infty
(u/2)^{2p+l} K_p^{(l)}(\alpha).
\end{equation}

As it comes, the $K_p^{(l)}(\alpha)$ is expressed in terms of Hermite
polynomials multiplied by finite powers of $\alpha$. It is tricky to
remove this dependence using the recursion formula on Hermite polynomials:
\begin{equation}
\label{eq:recforherm}
2 x H_n(x) = H_{n+1}(x) + 2n H_{n-1}(x).
\end{equation}

For small values of $l$, it appears that the $K_p^{(l)}$ can be put under
the form:
\begin{equation}
\label{eq:formK}
K_p^{(l)}(x) = \frac{l!}{p!} \sum_{r=0}^l \frac{(p+r)!}{r! (l-r)!}
\frac{H_{2p+2r+1}(x)}{(2p+2r+1)!}.
\end{equation}

What remains to do is to prove the general formula valid for any $l$. This
can be done by induction, using the well known recursion relation
\begin{equation}
\label{eq:recformil}
i_{l+1}(z) = i_{l-1}(z) - \frac{2l+1}{z}i_l(z).
\end{equation}

To get the final result we come back to the variable $z$ instead of $u$.
Thus, the series expansion of important use is given by
\begin{equation}
\label{eq:develseril} e^{-az^2} i_l(bz) = \frac{1}{2 \alpha ^{l+1}}
\sum_{p=0}^\infty \left(\sqrt{a}z\right)^{2p+l}K_p^{(l)}(\alpha); \,
\alpha = \frac{b}{2\sqrt{a}}
\end{equation}
with the $K_p^{(l)}$ function defined by (\ref{eq:formK}).

Of some interest is also the series expansion of Hermite polynomials appearing
in the $K_p^{(l)}(x)$ functions; explicitly
\begin{equation}
\label{eq:expHermit}
\frac{H_{2n+1}(x)}{(2n+1)!} = \sum_{r=0}^n (-1)^{n+r} \frac{(2x)^{2r+1}}
{(2r+1)! (n-r)!}.
\end{equation}

\section{Integral over angular variables}
\label{sec:intangvar}
In the process of calculation of the matrix elements in terms of those
concerned by the generating functions, we are faced to calculate the
following integral
\begin{equation}\begin{split}
\label{eq:intYLM}
&\mathcal{I}^{L'M',LM,n}_{\lambda \mu}(a',a) = \\
&\int d \hat{\bm e} d\hat{\bm e'}Y_{L'M'}^\ast(\hat{\bm e'})
Y_{LM}(\hat{\bm e})\mathcal{Y}_{\lambda \mu}(a \bm e + a' \bm e')
(\bm e \cdot \bm e')^n\end{split}
\end{equation}
where $\bm e$ and $\bm e'$ are unit vectors ($| \bm e | = 1 = | \bm e' |$)
and $\mathcal{Y}_{\lambda \mu}(\bm r) = r^\lambda Y_{\lambda \mu}
(\hat{\bm r})$ is a solid harmonic. The integration is done on angular
coordinates of $\bm e$ and $\bm e'$.

Because of rotational properties, we guess that this integral is proportional
to a Clebsch-Gordan coefficient. Thus, let us define a reduced matrix element
$\mathcal{I}^{L',L,n}_{\lambda }(a',a)$ through the usual form ($\widehat{L}
=\sqrt{2L+1}$)
\begin{equation}
\label{eq:elYLMred}
\mathcal{I}^{L'M',LM,n}_{\lambda \mu}(a',a) = \frac{\left\langle L M \lambda \mu
| L'M' \right\rangle}{\widehat{L'}}\mathcal{I}^{L',L,n}_{\lambda }(a',a).
\end{equation}

The evaluation of this quantity is based on three fundamental formulae
\begin{enumerate}
    \item
\begin{equation}
\label{eq:devYsol}
\mathcal{Y}_{\lambda \mu}(a \bm e + a' \bm e')= \sum_{l=0}^\lambda
Z_l^\lambda a^l a'^{\lambda - l}\left[Y_l(\hat{\bm e}) Y_{\lambda-l}
(\hat{\bm e'})\right]_{\lambda \mu},
\end{equation}
where the geometrical coefficient $Z_l^\lambda$ takes the value
\begin{equation}
\label{eq:coefZ} Z_l^\lambda = \sqrt{\frac{4\pi(2 \lambda
+1)!}{(2l+1)!(2\lambda - 2l +1)!}}.
\end{equation}
Note that in obtaining (\ref{eq:devYsol}), we used the fact that $\bm e$ and
$\bm e'$ are unit vectors. Note also the symmetry of the $Z$ coefficient:
$Z_l^\lambda=Z_{\lambda - l}^\lambda$.
    \item
\begin{equation}\begin{split}
\label{eq:devprodscalee}
(\bm e \cdot& \bm e')^n = \\
&\sum\limits_{k,p \geq 0, 2k+p =n} B_{k\,p} (-1)^p \sqrt{2p+1}
\left[Y_p(\hat{\bm e}) Y_p(\hat{\bm e'})\right]_{00}\end{split}
\end{equation}
which is valid for unit vectors $\bm e$ and $\bm e'$. The
geometrical coefficient $B_{k\,p}$ is defined through
(\ref{eq:BKL}). One must always have $k$ integer $\geq 0$ and  the
index $p$ must have the same parity than $n$ in
(\ref{eq:devprodscalee}).
    \item
\begin{equation}
\label{eq:contYl}
\left[Y_{l_1}(\hat{\bm e}) Y_{l_2}(\hat{\bm e})\right]_{lm} =
\frac{\widehat{l_1}\,\widehat{l_2}}{\sqrt{4 \pi} \, \widehat{l}} \left\langle
l_1\;0\;l_2\;0\vert l \; 0\right\rangle Y_{lm}(\hat{\bm e}).
\end{equation}
\end{enumerate}

Inserting those relations in the searched integral leads, after some
manipulations, to the final result:
\begin{equation}\begin{split}
\label{eq:finresint} \mathcal{I}^{L',L,n}_{\lambda}(a',a)& =
\frac{\widehat{\lambda}}{4 \pi} \sum\limits_{l=0}^\lambda (-1)^l \,
\widehat{l} \, \widehat{\lambda - l} \,
Z_l^\lambda a^l a'^{\lambda - l}  \\
& \times \sum\limits_{k,p \geq 0 \atop 2k+p =n} B_{k\,p} \; (2p+1)
\left\langle l\;0\;p\;0\vert L \; 0
\right\rangle\\
&\times  \left\langle \lambda -l\;0\;p\;0\vert L' \; 0\right\rangle
\begin{Bmatrix}
\lambda -l & l & \lambda \\
L & L' & p
\end{Bmatrix} \end{split}
\end{equation}
including usual Clebsch-Gordan and 6J coefficients.

For the tensor operator, we need this integral with the peculiar
value $\lambda=2$. Using the symmetry properties of Clebsch-Gordan
coefficients, it is easy to check that $L$ and $L'$ must have the
same parity, and, because of the angular momentum coupling, this
means that $L'=L \pm 2$ or $L' = L$. Inserting in
(\ref{eq:finresint}) the special values for the Clebsch-Gordan and
6J coefficients, one gets the value of the integral in this case
\begin{subequations}\label{eq:reslamb2}
\begin{equation}\begin{split}
 \mathcal{I}^{L \pm 2,L,n}_{2}(a',a)& =
\sqrt{\frac{15 L_m (L_m+1)}{8 \pi
(2L_m +1)}}\\
\times & \sum\limits_{l=0}^2 \sum\limits_{k \geq 0} 2^{\bar{l}}
B_{k\,(L_i+l)} a_{\pm}^l {a'}_{\pm}^{2-l} \delta_{2k+L_i,n-l},
\end{split}
\end{equation}
\begin{equation}\begin{split}
\mathcal{I}^{L,L,n}_{2}&(a',a) = - \sqrt{\frac{5 L(L+1)(2L+1)}{4 \pi
(2L-1)(2L+3)}} \\
\times & \sum\limits_{l=0}^2 \sum\limits_{k \geq 0}
\frac{4k+2L+2+\bar{l}} {2^{1-\bar{l}}(2k+L+1-\bar{l})} B_{k\,L} a^l
{a'}^{2-l} \delta_{2k+L,n+\bar{l}},\end{split}
\end{equation}
\end{subequations}
with $L_m,L_i,\bar{l}$ defined in (\ref{eq:Lmi}) and the quantities
\begin{equation}
\label{eq:defLmbarl}
(a_+,a'_+) =(a,a'); \; (a_-,a'_-) = (a',a).
\end{equation}

For the $LS$ operator, we need such an integral with $i (\bm e \times
\bm e')_\mu$ instead of $\mathcal{Y}_{\lambda \mu}(a \bm e + a' \bm e')$. Since
$i (\bm e \times \bm e')_\mu$=$\frac{4\pi \sqrt{2}}{3}[\mathcal{Y}_1(\bm e)
\mathcal{Y}_1(\bm e')]_{1 \mu}$, most of the calculations needed before can
be applied as well. One finds
\begin{equation}\begin{split}
\label{eq:prodveceep} &\int d \hat{\bm e} d\hat{\bm
e'}Y_{L'M'}^\ast(\hat{\bm e'}) Y_{LM}(\hat{\bm e}) [i (\bm e \times
\bm e')_\mu]
(\bm e \cdot \bm e')^n = \\
&\delta_{LL'}\frac{\sqrt{L(L+1)}}{n+1} \left\langle L\;M\;1\;\mu
\vert L' \; M' \right\rangle \sum\limits_{k \geq 0} B_{k\,L} \;
\delta_{2k+L,n+1}.\end{split}
\end{equation}

\section{Spin reduced elements for 3-body systems}
\label{sec:spinredel}
In this section we present the spin reduced matrix elements for the 3-body
problem that appear for the most important spin dependent operators. In some
papers we can find them but most of the time they are given for spin $1/2$
particles (nucleons or quarks). Here we have in mind the general 3-body
problem (in particular one can consider hybrid states including gluons with
spin 1, or pions with spin 0). The spin for the particle $i$ is denoted
$\bm s_i$.

The spin function for the the 3-body problem is chosen as
\begin{equation}
\label{eq:fospin3c}
\left| \chi_S (1,2,3) \right\rangle = \left| [(s_1 s_2)_{S_{12}}s_3]_S
\right\rangle
\end{equation}
where $S_{12}$ is the partial coupling of the $(1-2)$ pair.

We want to calculate the reduced matrix elements $\left\langle \chi'_{S'}
|| \hat{O} || \chi_S \right\rangle $ for the interesting spin operators
$\hat{O}$. In practice, we will consider
\begin{itemize}
    \item one-body spin operators of the form $\hat{O}_i (1,2,3)= \hat{O}_i(i)
\otimes \hat{1}(j) \otimes \hat{1}(k)$, where $\hat{1}(j)$ is the unit
operator for particle $j$ and where $\hat{O}_i(i)$ concerns the particle
$i$ only and has a given tensorial character.
    \item two-body spin operators of the form $\hat{O}_{ij} (1,2,3) =
\hat{O}_{ij}(i,j) \otimes \hat{1}(k)$. The operator $\hat{O}_{ij}(i,j)$
concerns the pair $(ij)$ and results itself from the coupling $[\hat{O}_i(i)
\otimes \hat{O}_j(j)]$.
\end{itemize}

To obtain the searched matrix elements is just a matter of Racah recoupling.
Very few formulae are indeed necessary.
The most important one concerns the case when the operator is the tensor
product of two operators acting on two distinct subsystems. Explicitly
($\hat{J}= \sqrt{2J+1}$)
\begin{equation}\begin{split}
\label{eq:elmatprodtens}
& \left\langle (j'_1 j'_2)_{J'} || [O_1^{k_1} \otimes O_2^{k_2}]_k ||
(j_1 j_2)_{J} \right\rangle = \\
&\hat{J'}\; \hat{J} \; \hat{k} \begin{Bmatrix} j_1 & j_2 & J \\ k_1
& k_2 & k \\ j'_1 & j'_2 & J' \end{Bmatrix} \left\langle j'_1 ||
O_1^{k_1} || j_1 \right\rangle \left\langle j'_2 || O_2^{k_2} || j_2
\right\rangle. \end{split}
\end{equation}

Very often, we are in the special case where $O_2^{k_2} = \hat{1}(2)$; then
one uses the special value
\begin{equation}
\label{eq:redmatunit}
\left\langle j' || \hat{1} || j \right\rangle = \hat{j} \; \delta_{j'j}
\end{equation}
to get a simplified relation
\begin{equation}\begin{split}
\label{eq:elmatprodtenspec}
 \bigg\langle (j'_1 j'_2)_{J'}&  || O_1^{k_1} ||(j_1 j_2)_{J} \bigg\rangle =
\delta_{j'_2 j_2} \left\langle j'_1 || O_1^{k_1} || j_1 \right\rangle \\
\times &(-1)^{j'_1+j'_2+J+k_1} \; \hat{J'} \; \hat{J}
\begin{Bmatrix} j'_1 & k_1 & j_1 \\ J & j_2 & J' \end{Bmatrix}.
\end{split}
\end{equation}

Of some utility is also the following formula
\begin{equation}
\label{eq:elmatredJ}
\left\langle j' || \hat{J} || j \right\rangle = \hat{j} \; \sqrt{j(j+1)}
\; \delta_{j'j}.
\end{equation}

With these tools, the calculation of the various reduced matrix elements is
just a matter of algebraic calculus.

When we consider the spin-orbit correction of a confining QCD potential,
the spin operators that intervene are simply the one-body $\bm s_i$
operators. Application of the above formulae leads to
\begin{eqnarray}\label{eq:spinopsi}
& \left\langle \chi'_{S'} || \bm s_1 || \chi_S \right\rangle =
(-1)^{s_1+s_2+s_3
+S_{12}+S'_{12}+S} \sqrt{s_1(s_1+1)}& \nonumber \\
& \times \widehat{s_1} \ \widehat{S_{12}} \ \widehat{S'_{12}} \
\widehat{S} \ \widehat{S'} \begin{Bmatrix} S'_{12} & 1 & S_{12} \\ S
& s_3 & S'
\end{Bmatrix} \ \begin{Bmatrix} s_1 & 1 & s_1 \\ S_{12} & s_2 & S'_{12}
\end{Bmatrix} &, \\
& \left\langle \chi'_{S'} || \bm s_2 || \chi_S \right\rangle = (-1)^{s_1+s_2-s_3
-S} \sqrt{s_2(s_2+1)}& \nonumber \\
& \times \widehat{s_2} \ \widehat{S_{12}} \ \widehat{S'_{12}} \
\widehat{S} \ \widehat{S'} \begin{Bmatrix} S'_{12} & 1 & S_{12} \\ S
& s_3 & S'
\end{Bmatrix} \ \begin{Bmatrix} s_2 & 1 & s_2 \\ S_{12} & s_1 & S'_{12}
\end{Bmatrix}, & \\
& \left\langle \chi'_{S'} || \bm s_3 || \chi_S \right\rangle = (-1)^{S'+S_{12}
+s_3+1} \delta_{S'_{12},S_{12}}& \nonumber \\
& \times \sqrt{s_3(s_3+1)} \ \widehat{s_3} \ \widehat{S} \
\widehat{S'} \
\begin{Bmatrix} s_3 & 1 & s_3 \\ S & S_{12} & S'
\end{Bmatrix}.&
\end{eqnarray}
For the two-body symmetric spin-orbit potential, the spin operator is
simply $\bm S_{ij} = \bm s_i + \bm s_j$. The reduced matrix elements for this
operator are easily obtained by addition of the corresponding $\bm s_i$ and
$\bm s_j$ elements as given previously.

For the two-body antisymmetric spin-orbit potential, the spin
operator is simply $\bm \Delta_{ij} = \bm s_i - \bm s_j$. The
reduced matrix elements for this operator are easily obtained by
difference of the corresponding $\bm s_i$ and $\bm s_j$ elements as
given previously.

Lastly for the two-body tensor operator the spin operator is $\mathcal{S}_{ij}$
with two possible expressions for the operator (see Eq. (\ref{eq:altformSij})).
After long but straightforward calculations, one gets

for $\mathcal{S}_{ij} =(\bm s_i \otimes \bm s_j)_2$
\begin{equation}\begin{split}
\label{eq:redmatspintens1} \big\langle \chi'_{S'}& || (\bm s_1
\otimes \bm s_2)_2|| \chi_S \big\rangle =
\widehat{s_1} \ \widehat{s_2}\sqrt{5s_1s_2(s_1+1)(s_2+1)}  \\
& \times (-1)^{S+S'_{12}+s_3+1}  \ \widehat{S_{12}} \
\widehat{S'_{12}} \
\widehat{S} \ \widehat{S'} \\
& \times  \begin{Bmatrix} S'_{12} & 2 & S_{12} \\ S & s_3 & S'
\end{Bmatrix} \ \begin{Bmatrix} s_1 & s_2 & S_{12} \\ 1 & 1 & 2 \\ s_1 & s_2 &
 S'_{12} \end{Bmatrix} ,\\
\end{split}\end{equation}

\begin{equation}\begin{split}
\label{eq:redmatspintens2} \big\langle \chi'_{S'} & || (\bm s_2
\otimes \bm s_3)_2|| \chi_S \big\rangle =
\sqrt{5s_2s_3(s_2+1)(s_3+1)}  \\
&\times \widehat{s_2} \ \widehat{s_3} \ \widehat{S_{12}} \
\widehat{S'_{12}} \ \widehat{S} \ \widehat{S'} \sum_{k,k'}
(-1)^{S+k+s_1+1} \\ &\times (2k+1)(2k'+1)
\begin{Bmatrix} s_1 & s_2 & S'_{12} \\ s_3 & S' & k' \end{Bmatrix}
\begin{Bmatrix} s_1 & s_2 & S_{12} \\ s_3 & S & k \end{Bmatrix} \\
&\times \begin{Bmatrix} k' & 2 & k \\ S & s_1 & S' \end{Bmatrix}
\begin{Bmatrix} s_2 & s_3 & k \\ 1 & 1 & 2 \\ s_2 & s_3 & k' \end{Bmatrix}.
\end{split}\end{equation}

for $\mathcal{S}_{ij} =(\bm S_{ij} \otimes \bm S_{ij})_2$
\begin{equation}\begin{split}
\label{eq:redmatspintens3} \big\langle \chi'_{S'}& || (\bm S_{12}
\otimes \bm S_{12})_2|| \chi_S \big\rangle = \delta_{S_{12},S'_{12}} \sqrt{5}
S_{12} (S_{12}+1) (2S_{12}+1)\\
& \times (-1)^{S+s_3+1-S_{12}}  \ \widehat{S} \ \widehat{S'} \\
& \times  \begin{Bmatrix} S_{12} & 2 & S_{12} \\ S & s_3 & S'
\end{Bmatrix} \ \begin{Bmatrix} 1 & 1 & 2 \\ S_{12} & S_{12} &
 S_{12} \end{Bmatrix} ,\\
\end{split}\end{equation}

\begin{equation}\begin{split}
\label{eq:redmatspintens4} \big\langle \chi'_{S'} & || (\bm S_{23}
\otimes \bm S_{23})_2|| \chi_S \big\rangle = \sqrt{5}
\ \widehat{S_{12}} \ \widehat{S'_{12}} \ \widehat{S} \ \widehat{S'} \\
&\times \sum_{k} (-1)^{S-k+s_3+1} k(k+1)(2k+1)^2 \\
&\times \begin{Bmatrix} 1 & 1 & 2 \\ k & k & k \end{Bmatrix}
\begin{Bmatrix} k & 2 & k \\ S & s_3 & S' \end{Bmatrix} \\
&\times \begin{Bmatrix} s_1 & s_2 & S_{12} \\ s_3 & S & k \end{Bmatrix}
\begin{Bmatrix} s_1 & s_2 & S'_{12} \\ s_3 & S' & k \end{Bmatrix}.
\end{split}\end{equation}

The matrix element $\left\langle \chi'_{S'} || \mathcal{S}_{13}||
\chi_S \right\rangle$ is obtained from the element $\left\langle
\chi'_{S'} || \mathcal{S}_{23}|| \chi_S \right\rangle$ with the
interchange $1 \leftrightarrow 2$ and adding a phase $(-1)^{S_{12} -
S'_{12}}$.

\section{Summary of previous known formulae}
\label{sec:formconnu}

This section is devoted to the presentation of formulae that are not new
since they have been derived previously either in SV or in SBM. Nevertheless
we find convenient to give a brief summary of them for two main reasons:
\begin{itemize}
    \item to achieve some unity in the paper since with the bulk of formulae
given in this section and in the rest of this work, the reader has all
the necessary tools for solving the few-problem with the most common
types of potentials.
    \item we want to show that whatever the dynamical quantities that are
considered, they all rely on the same geometrical coefficients and a few
of universal geometrical functions.
\end{itemize}

All the dynamical parameters that appear in this section have been defined
previously in the paper.

\subsection{Overlap}
\label{subsec:Overlap}

The overlap between basis states
$\mathcal{N}_{K'KL}$ = $\left\langle \psi_{K'LM}(u',A') |
\psi_{KLM}(u,A) \right\rangle$ is a crucial ingredient in the
equation of motion. Since the basis states are non orthogonal, there
is no reason that such an element is diagonal.

As shown in SV, the overlap is
 [SV, (A.6) page 248]

\begin{eqnarray}
\label{eq:recgen} \mathcal{N}_{K'KL}=\frac{(2K'+L)! \:
(2K+L)!}{B_{K'L}B_{KL}}\left( \frac{\pi^N}{\det B}\right)^{3/2}
\\\nonumber \times\!\!\! \sum_{k=0}^{\min (K,K')}
B_{k\,L}
\frac{q^{K-k}}{(K-k)!}\frac{q'^{K'-k}}{(K'-k)!}\frac{\rho^{2k+L}}{(2k+L)!}
\end{eqnarray}

For the important peculiar case $K' = K = 0$, this formula
simplifies a lot and we are left with [SV (A.7) page 249]:

\begin{equation}
\label{eq:recpart} \mathcal{N}_{00L}= \mathcal{N}_L =
\frac{(2L+1)!!}{4 \pi}\left( \frac{\pi^N}{\det B}\right)^{3/2}
\rho^L.
\end{equation}

\subsection{Non relativistic kinetic energy}
\label{sec:Tnr}

The intrinsic kinetic energy operator $T_{NR}$ can be cast under the form

\begin{equation}
\label{eq:Tnr1}
T_{NR}  = \frac{1}{2} \sum_{i,j=1}^N \Lambda_{ij} \bm
\pi_i \cdot \bm \pi_j = \frac{1}{2} \tilde{\bm \pi} \cdot \Lambda
\bm \pi
\end{equation}

The final result for this operator is [see SV, (A.10) page 250]
\begin{equation}\begin{split}
\label{eq:Tnrgen} \big\langle & \psi_{K'LM}(u',A') | \tilde{\bm\pi}
\cdot \Lambda \bm \pi | \psi_{KLM}(u,A) \big\rangle
= \\
\times&\frac{(2K'+L)! \: (2K+L)!}{B_{K'L}B_{KL}}
\left(\frac{\pi^N}{\det
B}\right)^{3/2}\ \  \sum_{k=0}^{\min (K,K')} B_{k\,L}  \\
\times &\left[ Rqq'\rho +
P(K-k)q'\rho + P'(K'-k)q\rho + Q(2k+L)qq'\right]  \\
\times& \frac{q^{K-k-1}}{(K-k)!} \frac{q'^{K'-k-1}}{(K'-k)!}
\frac{\rho^{2k+L-1}}{(2k+L)!}\end{split}
\end{equation}
with the numbers $P,P',Q,R$ defined by
\begin{eqnarray}
\label{eq:PPQR} P=-\tilde{u}B^{-1}A'\Lambda A'B^{-1}u&;& \quad
P'=-\tilde{u'}B^{-1}A\Lambda AB^{-1}u'; \nonumber \\
Q = 2\tilde{u'}B^{-1}A\Lambda A'B^{-1}u&;& \quad R = 6
\textrm{Tr}(AB^{-1}A' \Lambda).
\end{eqnarray}

Again, the formula for the special case $K = K' = 0$ is much simpler
\begin{equation}
\label{eq:Tnrpart} \left\langle \psi_{0LM}(u',A') | \tilde{\bm \pi}
\cdot \Lambda \bm \pi | \psi_{0LM}(u,A) \right\rangle =
\mathcal{N}_L \left(R + L \frac{Q}{\rho}\right).
\end{equation}

In SBM, we gave also the matrix elements for a semi-relativistic kinetic
energy operator but, since it is not of use in atomic and nuclear physics
we do not report it here. The interested reader will find it under
the formula (44) of SBM.

\subsection{Central potentials}
\label{subsec:centpot} In the most interesting cases, the potentials
appearing in the few-body problem are either one-body potentials or
two-body potentials. With the same arguments than those developed in
the second section, the most general form of the central potential
is a sum of terms like $V(| \tilde{w} \bm x |)$.

The expression given in SV looks similar to [SV, (A.128) page 282]:
\begin{equation}
\label{eq:Vcentgen}
\begin{split}
\langle &\psi_{K'LM}(u',A') | V(|\tilde{w} \bm x|) | \psi_{KLM}(u,A)
\rangle =\\& \left(\frac{\pi^N}{\det B}\right)^{3/2}\frac{(2K'+L)!
\: (2K+L)!}{B_{K'L}B_{KL}}
\sum_{n=0}^{K+K'+L} \frac{J(n,c)}{c^n}\\
&\times   \sum_{k=0}^{\min (K,K')}  B_{k\,L}
F^n_{K-k,K'-k,2k+L}(q,q',\rho,\gamma,\gamma')
\end{split}
\end{equation}
which, for the peculiar case $K=K'=0$ reduces to [SV, (A.130) page
282]
\begin{equation}
\label{eq:Vcentpart} \left\langle \psi_{0LM} | V(|\tilde{w} \bm x|)
| \psi_{0LM} \right\rangle = \mathcal{N}_L \: L! \sum_{n=0}^L
\frac{J(n,c)}{(L-n)!} \left(\frac{\gamma \gamma'}{\rho c}\right)^n.
\end{equation}

Expressed in terms of the $\mathcal{F}$ integrals, the value can be
found in SBM [SBM, (C14) page 10]
\begin{equation}\label{eq:Vcentgen2}\begin{split}
 \big\langle &\psi_{K'LM}(u',A') | V(|\tilde{w}
\bm x|) | \psi_{KLM}(u,A) \big\rangle =  \\
&\frac{(2K'+L)! \: (2K+L)!}{\sqrt{2\pi} 2^{K+K'-1} B_{K'L}B_{KL}}
  \left(\frac{c \pi^N}{\det B}\right)^{3/2} \frac{\gamma^{2K+L}
{\gamma'}^{2K'+L}}{(-c)^{K+K'+L}} \\
\times&\sum_{n=0}^{K+K'+L} \frac{1}{(2n+1)!} (-2c)^n
\mathcal{F}_V(2n+2,c/2) \\
\times&\sum_{k=0}^{\min(K,K')} 4^k B_{kL} H^{K,K',L}_{n,k} \left(
\frac{2q \gamma'}{\rho \gamma},\frac{2q' \gamma}{\rho \gamma'} ,
\frac{\rho c}{\gamma \gamma'} \right)\end{split}
\end{equation}
while, for the special case $K = K' = 0$, this formula reduces to
[SBM, (C17) page 10]
\begin{eqnarray}
\label{eq:Vcentpart2}
& \left\langle \psi_{0LM}(u',A') | V(|\tilde{w}
\bm x|) | \psi_{0LM}(u,A) \right\rangle = \mathcal{N}_L \: 2c
\sqrt{\frac{c}{2 \pi}} \: L! & \nonumber \\
&\times \sum_{n=0}^L \frac{(2c)^n
\mathcal{F}_V(2n+2,c/2)}{(2n+1)!(L-n)!} \left(\frac{\gamma
\gamma'}{\rho c}\right)^n \left(1 - \frac{\gamma \gamma'} {\rho
c}\right)^{L-n}.&
\end{eqnarray}

Following the philosophy proposed in SBM, an alternative formula exist
for the central matrix elements [SBM, (16) page 3]
\begin{equation}\begin{split}
\label{eq:Vcentgen3} \big\langle &\psi_{K'LM}(u',A') | V(|\tilde{w}
\bm x|) | \psi_{KLM}(u,A) \big\rangle = \left(\frac{\alpha
\pi^N}{\det B}\right)^{\frac{3}{2}}  \\
\times &\frac{(2K'+L)! \: (2K+L)!}{B_{K'L}B_{KL}}
\sum_{n=0}^{K+K'+L}
\left(\frac{\alpha}{2c}\right)^n J(n,\alpha,c) \\
\times & \sum_{k=0}^{\min (K,K')}  \frac{2^{2k+L}}{(2k+L)!} B_{k\:L}
F^{K,K',L}_{n,k}(\bar{q},\bar{q}\,',\gamma,\gamma'). \end{split}
\end{equation}
which, in the special case $K=K'=0$ reduces to [SBM, (21) page 4]
\begin{equation}
\label{{eq:Vcentpart2}} \left\langle \psi_{0LM}| V(|\tilde{w} \bm
x|) | \psi_{0LM} \right\rangle = \mathcal{N}_L \: L!
\left(\frac{\alpha}{1-\alpha} \right)^L \alpha^{3/2}
J(L,\alpha,c).
\end{equation}

Using rather the $\mathcal{F}$ integrals gives another formulation
[SBM, (22) page 4]
\begin{equation}
\label{eq:Vcentgen4} \begin{split} &\left\langle
\psi_{K'LM}(u',A') V(|\tilde{w} \bm x|) | \psi_{KLM}(u,A)
\right\rangle =\\& \frac{4 (2K'+L)! \: (2K+L)!}{\sqrt{\pi}
B_{K'L}B_{KL}} \left(\frac{c \pi^N}{2 \det B}\right)^{3/2} \\ &
\times \sum_{n=0}^{K+K'+L} \frac{1}{(2n+1)!}\mathcal{F}_V(2n+2,c/2) \\
&\times \sum_{k=0}^{\min (K,K')}\frac{2^{2k+L}}{(2k+L)!} B_{k\:L}
P^{K,K',L}_{n,k}(\bar{q},\bar{q}\,',\gamma,\gamma',c),
\end{split}
\end{equation}
while the special case $K=K'=0$ gives again the formula
(\ref{eq:Vcentpart2}) presented before.

Comparing the formulae given in this section with the new ones proposed
in the paper shows the tight similarities that exist between all the
types of potential.

\end{document}